\documentclass[
reprint,
amsmath,amssymb,
aps,
nofootinbib
]{revtex4-2}

\usepackage{graphicx}
\usepackage{dcolumn}
\usepackage{bm}
\usepackage{braket}
\usepackage{siunitx}
\usepackage{float}
\usepackage[bottom]{footmisc}
\usepackage{subfigure}
\usepackage{tabularx}
\usepackage{makecell}
\usepackage{footnote}
\usepackage{calrsfs}

\newcommand{\La}{\mathcal{L}}

\begin{document}

\title{An Intense, Continuous Cold Atom Source}
\author{William Huntington, Jeremy Glick, Michael Borysow, and Daniel J. Heinzen }
\affiliation{%
 Dept. of Physics, The University of Texas, Austin, TX 78712
 }

\date{\today}

\begin{abstract}
We demonstrate an intense, continuous cold atom beam generated via post nozzle seeding of a supersonic helium jet with $^7$Li atoms. The nozzle is cooled to about 4.4 K to reduce the forward velocity of the atoms. The atomic beam is brought to a focus 175 cm from the nozzle by a 10 cm bore diameter magnetic hexapole lens. Absorption and fluorescence imaging of the focus show a flux of \SI{2.3(4)e12}{} atoms/s,  brightness of \SI{4.1(7)e19}{} 
$\textrm{m}^{-2}\textrm{s}^{-1}\textrm{sr}^{-1}$, forward velocity of 211(2) m/s, and longitudinal temperature of 7(3) mK. Results agree with a Monte Carlo simulation of the seeding dynamics and a particle tracing simulation of the atom lens. We project that 10 times higher flux would be possible with improved vacuum system design. Our method should provide a useful high-brightness source for atom-optical and other atomic and molecular physics applications.

\end{abstract}
\maketitle

\section{Introduction}

Cold atomic and molecular beams play a crucial role in experimental physics. They make possible the efficient loading of atom traps and studies of quantum gases   \cite{Pethick_2008_BEC_book, Inguscio_2016_quantum_matter}. Cold atomic beams and fountains are  widely used for precision measurement experiments, including searches for time-reversal symmetry violation,  atomic clocks, and atom interferometers  \cite{Roberts_2015_P_and_T_violation, Wynands_2005_fountain_clock, Bong_2019_atom_interferometer_review, Tino_2021_gravity_atom_interferometry}. Cold, slow beams of molecules are of interest for studies of ultracold molecular collisions, ultracold chemistry, molecular quantum gases, and precision measurement applications  \cite{Jankunas_2015_cold_molecular_beams, Dulieu_2018_cold_chemistry, Heazlewood_2021_cold_chemistry_review, Hudson_2011_YbF_edm, Andreev_2018_ACME_edm}.  

Laser cooling has been the dominant method for producing cold atomic beams \cite{Phillips_1982_Zeeman_slower, Hau_2005_Zeeman_slower, Lu__1996_LVIS, Lee_1996_pryamid_MOT, Dieckmann_1988_2D_mot}. However, laser cooling is not easily applied to all atoms, and is generally more difficult for molecules than for atoms. Nevertheless, it has been demonstrated that a variety of molecules can be laser-cooled, and applications of such laser-cooled molecules are being pursued \cite{Fitch_2021_laser_cooled_molecules}.

Seeded, supersonic rare gas jets provide another route to producing cold beams, in which low temperatures are produced by the adiabatic expansion of the carrier gas. These have long been used to cool the rotational and vibrational degrees of freedom of molecular beams \cite{Smalley_1977_molecular_spectroscopy}. Rotational and translational temperatures in the range from one to a few Kelvin are generally achievable \cite{Smalley_1977_molecular_spectroscopy, Tarbutt_2002_cold_YbF_beam, Yan_2013_short_pulse_molecular_beam, Aggarwal_2021_Seeded_SrF_beam}, and temperatures as low as 0.2 K have been reported \cite{Hillenkamp_2003_condensation_limited_cooling, Melin_2019_1D_Stern_Gerlach}. Vibrational temperatures are generally much higher than rotational temperatures \cite{Smalley_1977_molecular_spectroscopy, Tarbutt_2002_cold_YbF_beam}. 
Such seeded jets often have large forward velocities, particularly if helium is used as the carrier gas. However, the velocity of such beams can be reduced with switched electric \cite{Bethlem_1999_electric_slower} or magnetic \cite{Narevicius_2007_pulsed_magnetic_slower} slowers, and this has led to a number of cold molecule applications \cite{Jankunas_2015_cold_molecular_beams, Heazlewood_2021_cold_chemistry_review}.

The buffer gas beam \cite{Hutzler_2012_buffer_gas_beam} provides another method for producing cold atomic and molecular beams. With this method, atoms or molecules are seeded into a carrier gas contained in a cryogenically cooled cell. Seeding is most commonly produced with pulsed laser ablation of a solid target. The carrier gas escapes through an orifice or other exit structure into a vacuum chamber and forms a beam containing the seeded atoms.  The resulting beam velocities, typically 40 to 150 m/s, are low due to the low temperature of the carrier gas. The buffer gas cools as it expands, but not to the same degree as a fully supersonic expansion. Beam temperatures are typically in the range of one to a few Kelvin. The buffer gas method has the advantage of very wide applicability, including to species that cannot be laser-cooled. 

Due to this wide and important range of applications, methods to further increase the flux, brightness, and range of species of molecular beams are of interest. In the work reported here, we explore a method to produce an intense, cold beam of $^7\textrm{Li}$ atoms with post-nozzle seeding of a cryogenic, supersonic $^4\textrm{He}$ jet, followed by extraction of the seeded atoms with a magnetic lens. Both the $^4\textrm{He}$ jet and $^7\textrm{Li}$ source are continuous rather than pulsed. Our method shares some features of the buffer gas method, including a relatively low forward velocity and applicability to a wide range of paramagnetic species without use of laser cooling. However, by using a supersonic expansion, we are able to take full advantage of the adiabatic cooling of the expanding helium, which can reach 1 mK temperature in the moving frame \cite{Wang_1988_1_mK_jet}. This allows us to reach very low $^7\textrm{Li}$ temperatures. Further, because the $^7\textrm{Li}$ beam is mono-energetic, we are able to magnetically focus it without suffering excessive chromatic aberration. Focusing of mono-energetic beams has been realized previously for metastable rare gas jets and laser-cooled beams \cite{Kaenders_1996_atom_lens, Chaustowsk_2007_magnetic_lens, Anciaux_2018_beam_brightening}, but not to our knowledge for seeded supersonic beams. This magnetic focusing enhances the intensity of the extracted $^7\textrm{Li}$ beam and makes possible its separation from the $^4$He beam for further use in a high vacuum application. We also optimize our source to produce a very high flux of $^7\textrm{Li}$ atoms.

Using this method, we have demonstrated a $^7\textrm{Li}$ beam with a flux of $2\times10^{12}$ atoms/s, a forward velocity of 210 m/s and a longitudinal temperature of 7 mK in the moving frame. To our knowledge this temperature is substantially lower than that of any previous seeded jet source. This combination of very low temperature, continuous output, modest forward velocity, magnetic extraction, and high flux distinguish our work from previous work on seeded supersonic jets and buffer gas sources. Also, as discussed below, we believe that at least ten times greater output flux should be achievable with improvements to the apparatus.

In section 2 of this article, we describe the design and theory of this apparatus and its components. In section 3 we describe our results of measurements of flux, temperature, brightness and other relevant features. In section 4, models of our experiment, including an original Monte-Carlo simulation of seeding dynamics, are discussed.

\begin{figure}[t]
    \centering
    \includegraphics[width=\columnwidth]{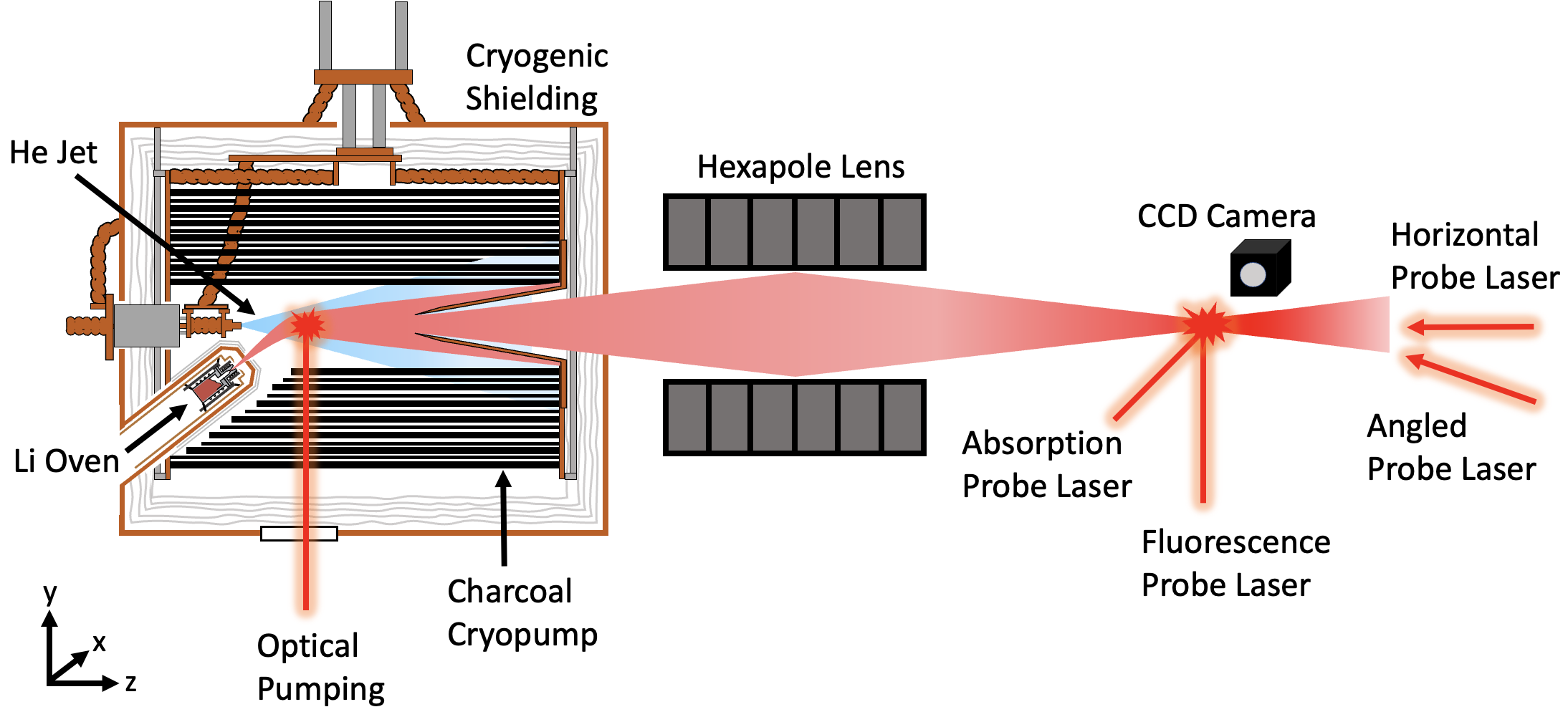}
    \caption{Overview of experimental apparatus. $^7$Li, produced from an oven, is seeded and thermalized into a supersonic helium jet. The jet is created by flowing helium gas through a nozzle inside of a cryogenic region that is cooled with a 2-stage pulse tube refrigerator. A charcoal cryo-adsorption pump is used to remove background helium while the $^7$Li is extracted via a skimmer. The extracted $^7$Li beam is then brought to a focus by a magnetic hexapole lens and is characterized using fluorescence and absorption spectroscopy.}
    \label{fig:Apparatus}
\end{figure}

\section{Apparatus}

\subsection{Design overview}

A diagram of our cold atomic beam source is shown in Fig.
\ref{fig:Apparatus}. $^4\textrm{He}$ gas is fed into a small cylindrical copper cell and cooled to a temperature of about 4.4 K by a cryo-refrigerator. The gas exits the chamber through a small hole and expands supersonically into a vacuum chamber. An effusive lithium beam produced from an oven beneath the nozzle is directed into the helium jet. A small fraction of the lithium atoms are entrained in the flow and are cooled to a relatively low temperature by collisions with the helium gas. The expanding gas is intercepted by a skimmer, which allows a central core of the beam into a following room temperature vacuum chamber. Helium that does not enter the skimmer is pumped away by a charcoal cryosorption pump \cite{Vasquez_1988_charcoal_pump}. 

The lithium beam enters a magnetic hexapole lens, which brings the lithium atoms to a focus by acting on their magnetic moment. In order to produce the highest $^7$Li flux at the focus, we optically pump the $^7$Li atoms into their $F=2,M=2$ magnetic sublevel just after the point of entrainment. This ensures that all atoms with the same trajectory into the lens experience the same focusing force. Helium atoms are not focused by the magnet, and therefore continue on ballistic trajectories until they hit a room temperature surface. These atoms are removed by a large diffusion pump below the magnet. 

The properties of our source at the focal point, discussed below, are the main result of this paper. 

\subsection{Supersonic Helium Jet}

We chose our design with the goal to produce the largest possible time-averaged flux of extracted lithium atoms at the lowest possible temperature. This led us to choose a continuous helium jet and source, in order to avoid possible negative effects of mismatch in the timing of pulsed helium or lithium sources. Further, we decided to operate the helium jet at the highest flux and brightness possible, on the assumption that this would also tend to maximize the achievable flux and brightness of the seeded atom beam. 

We generated our helium jet by flowing helium gas through a simple sonic nozzle consisting of a $200~\mathrm{\mu m}$ diameter hole drilled in the thin, flat output face of the cold copper cell.  Although it is possible that a shaped nozzle would produce better results, we chose this sonic nozzle jet for its well-understood properties \cite{miller_1988_free_jet}. Helium flow is metered by an MKS flow meter. Heat exchangers connected to the two stages of our refrigerator cool the gas so that it reaches a temperature $T_0$ in the range from 4.2 K to 4.6 K inside the copper cell. The number of atoms per second leaving the nozzle is \cite{miller_1988_free_jet}

\begin{equation}\label{Hefloweq}
    \dot{N}_{He} = 0.403 \frac{P_0}{k_B T_0} u_0 d^2
\end{equation}

\noindent where  $P_0$ is the stagnation (pre-nozzle) gas pressure, \textit{d} the nozzle diameter, and $u_0=\sqrt{2 k_B T_0/m_{He}}$, with $m_{He}$ the mass of a helium atom. 

In order to reach milliKelvin temperature, it is necessary to avoid the formation of helium clusters, since otherwise the heat of condensation is released into the expanding gas \cite{Hillenkamp_2003_condensation_limited_cooling}. The onset of cluster formation is determined by the Hagena parameter \cite{hagena_1987}

\begin{equation}\label{Hagenaeq}
    \Gamma^* = \kappa \frac{P_0 d^{0.85}}{T_0 ^{2.29}}
\end{equation}

\noindent where $\kappa$ is a species-specific condensation parameter. Over a fairly wide range of parameters and gas species, it is found that substantial cluster formation occurs only when $\Gamma^* > 300$. This scaling law has not been tested very near our parameter range, but He cluster formation in cryogenic jets with much smaller nozzles than ours has been studied \cite{STEPHENS_1983_cyrogenic_He_jet, BUCHENAU_1990_helium_cluster_beams, Bruch_2002_He_dimers_and_trimers}. In this work, some formation of dimers and trimers is observed with Hagena parameter as low as $\Gamma^* \approx 50$, but heating of the jet generally remains small up to $\Gamma^* \approx 300$. 

Helium has an unusually low $\kappa$ of $3.85 \: \mathrm{K^{2.29} \mu m^{-0.85} mbar^{-1}}$ \cite{helium_K, k_values}, due to the weakness of the He-He interaction, and to the fact that the helium dimer has only one bound state with an extremely small binding energy of about 1.1 mK \cite{Grisenti_2000_He_dimer}. This also produces an elastic collision cross-section that increases dramatically as the jet temperature falls \cite{Chrysos_2017_He_He_cross_section}. These factors allow helium jets to remain collisional to much larger distances and to cool to much lower temperatures than jets of other atoms.

The terminal forward velocity of the jet is $v_f = 1.58 \: u_0$ \cite{miller_1988_free_jet}. Ordinarily this is quite large (for instance 1750 m/s at room temperature), which is a downside of seeded supersonic jets. However our cryogenic nozzle produces a much smaller jet velocity. During the experiments we noticed that the nozzle temperature fluctuated due primarily to changes in the heat load of the helium flow, and this caused problematic variations of the jet velocity. To eliminate this problem, we stabilized the nozzle temperature with a heater and PID controller, typically at a value that produced a measured jet velocity of $v_f = 210 \: \textrm{m/s}$. The stabilized nozzle temperature was a few tenths of a degree above its value with no helium flow, and typically in the range from 4.2 K to 4.6 K. 

We designed our apparatus for a maximum flow rate of 300 sccm ($\dot{N}_{He} = 1.3 \times 10^{20}$ atoms/s), which is the limit set by the capacity of our charcoal cryosorption pump as discussed below. We chose the nozzle diameter $d = 200\:  \mathrm{\mu m}$ in order to maximize helium beam brightness without producing an excessively large Hagena parameter. For instance with $P_0 = 25 \: \mathrm{mbar}$ and $T_0 = 4.4 \: \mathrm{K}$, our nozzle produces a flow of 200 sccm and has a Hagena parameter $\Gamma^* = 290$. So far, we have not seen any clear evidence that our results are affected by helium cluster formation.

As the helium gas expands, its density drops and therefore its collision rate drops. This results in a region of continuum flow near the nozzle, a region of molecular flow far from the nozzle (assuming that no shocks form), and an intermediate flow region between these two \cite{miller_1988_free_jet, Pauly_2000_beams_book}. In the intermediate region, collisions are no longer frequent enough to maintain thermal equilibrium and the temperature for motions parallel and perpendicular to the nominal jet velocity become unequal. 

At distances $z \gtrsim 3d$, but still in the continuum flow region, the helium temperature \textit{T} falls with distance \textit{z} from the nozzle as \cite{Pauly_2000_beams_book}

\begin{equation}\label{adiabatic_cool}
    T(z) = 0.287~T_0 \left( \frac{z}{d} \right)^{-4/3}
\end{equation}

\noindent In practice, we found that defects on the inside surface of our nozzle hole could affect the measured temperature of the seeded $^7$Li. Presumably, this is evidence of an elevated helium temperature due to these defects. Polishing the nozzle with Simichrome to achieve a smooth surface remedied this issue. 

Adiabatic cooling as given by Eq. \ref{adiabatic_cool} extends only as far as the continuum flow region. Using the  known He-He cross section \cite{Chrysos_2017_He_He_cross_section} and density profile of our jet \cite{densityProfile}, we estimate that the helium mean-free path at $z = 4 \: \mathrm{cm}$ with a flow rate of 50 sccm is 2 mm, so that continuum flow conditions extend to at least $z = 4 \: \mathrm{cm}$.  Thus, according to Eq. \ref{adiabatic_cool}, we expect that our He gas should reach a temperature of 1 mK.  We also estimate only a few collisions remain for each helium atom at distances greater than $z = 10 \: \mathrm{cm}$.

A common feature of supersonic jets is the formation of shock fronts \cite{shockFronts, densityProfile}. These form as the supersonic flow, which travels faster than the local speed of sound, is unable to ``sense" downstream boundary conditions. This results in an over-expansion of the jet, which is compensated by the formation of shock fronts. These non-isentropic regions add entropy to the jet, heating it and potentially obstructing the extraction of the seeded species. When shocks are present, supersonic beam experiments use a skimmer \cite{miller_1988_free_jet} to penetrate through the shock front and extract the central part of the beam. 

In our experiment, we extract the beam from the cryogenic region with a skimmer located $16$ cm from the nozzle with a inlet diameter of $2.54$ cm, as shown in Fig. \ref{fig:cryogenics}. This is in the free molecular flow zone of our beam. We placed the skimmer at this point because we wanted to avoid skimmer interference effects that could occur for placement closer to the jet, and because we estimated that with the high speed of our charcoal pump, no shock would form out to this distance, even at our highest flow rate. Thus far, we have seen no experimental evidence of any negative effect of a shock front on beam extraction. 

The opening of our skimmer is chosen to easily pass Li atoms that are capturable by our magnetic lens, which lie within a half-angle of about 0.064 rad. Since this is a relatively small angle, it is important that entrainment is maximized along the centerline of the jet. To maximize seeding efficiency and alignment to the skimmer, our nozzle can be displaced and tilted with an adjustable bellows mount. 

\begin{figure}[t]
    \centering
    \includegraphics[width=\columnwidth]{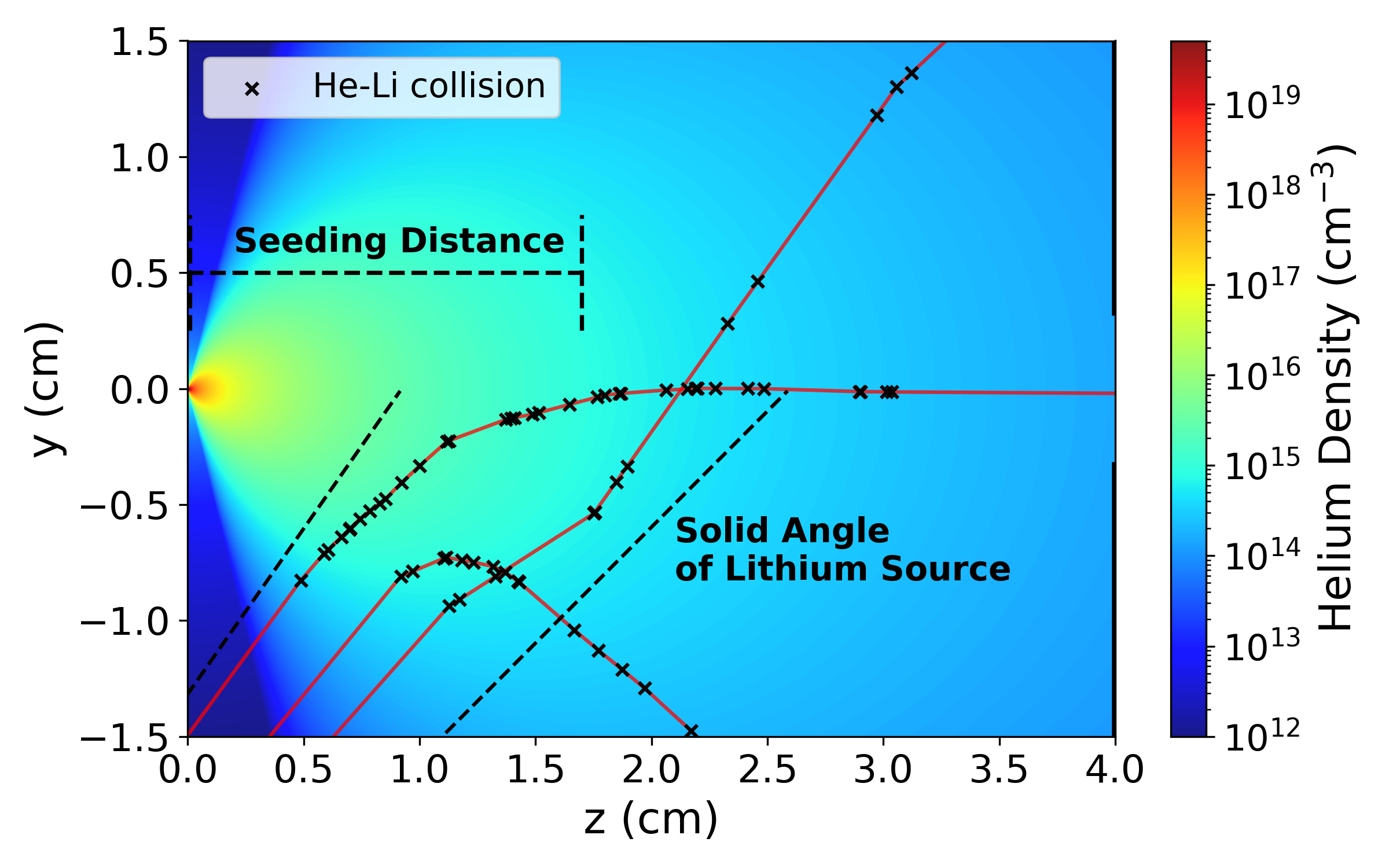}
    \caption{Example simulation results for $200$ sccm of He flow. Trajectories are shown along with the location of collisions denoted with an ``x". Three distinct type of particles are shown: one which is deflected by the jet, one seeded and within the capturable solid angle of the skimmer (half angle of 0.079 radians) and one which undergoes collisions but passes through the jet without fully thermalizing.}
    \label{seedingExample}
\end{figure}

In order to efficiently capture and extract seed atoms, the jet must be of intermediate transverse collisional thickness - not so low that many atoms pass through the jet, but not so high that many atoms are stopped before reaching the centerline of the jet. The transverse collisional thickness varies with distance $z$ as $1/z$, so it possible to increase or decrease the collisional thickness with a change in the seeding location. When estimating this location, it is important to take into account the energy dependence of the Li-He cross-section. Fortunately, this cross-section has been calculated to high accuracy \cite{Makrides_2020_CrossSection, Tiesinga_2021_CrossSection}. The first few Li-He collisions will have high relative energies, for which the cross-section is small. This means that the lithium atoms can easily penetrate into the jet even at locations where the helium jet is collisionally thick to itself. However after the first few collisions the relative collision energy drops and the cross-section increases very substantially. This can result in a pronounced increase in the collision rate, and cause the lithium atom to become entrained with the He flow and to approach the He jet temperature. 

Example trajectories from a Monte-Carlo simulation, discussed in further detail in the modeling section, are given in Fig. \ref{seedingExample}. The seeding geometry is shown as well as the trajectory and collision locations for a seeded particle, deflected particle, and one which passes through the jet without fully thermalizing. The simulation shows that for 200 sccm helium flow rate, lithium atoms have an appreciable chance of becoming entrained near the centerline of the helium flow if they are aimed at a point about 1 to 2 cm away from the helium nozzle. For lower helium flow rates, the optimal seeding distance is smaller.

\subsection{Cryogenics and Vacuum}

\begin{figure}[t]
    \centering
    \includegraphics[width=\columnwidth]{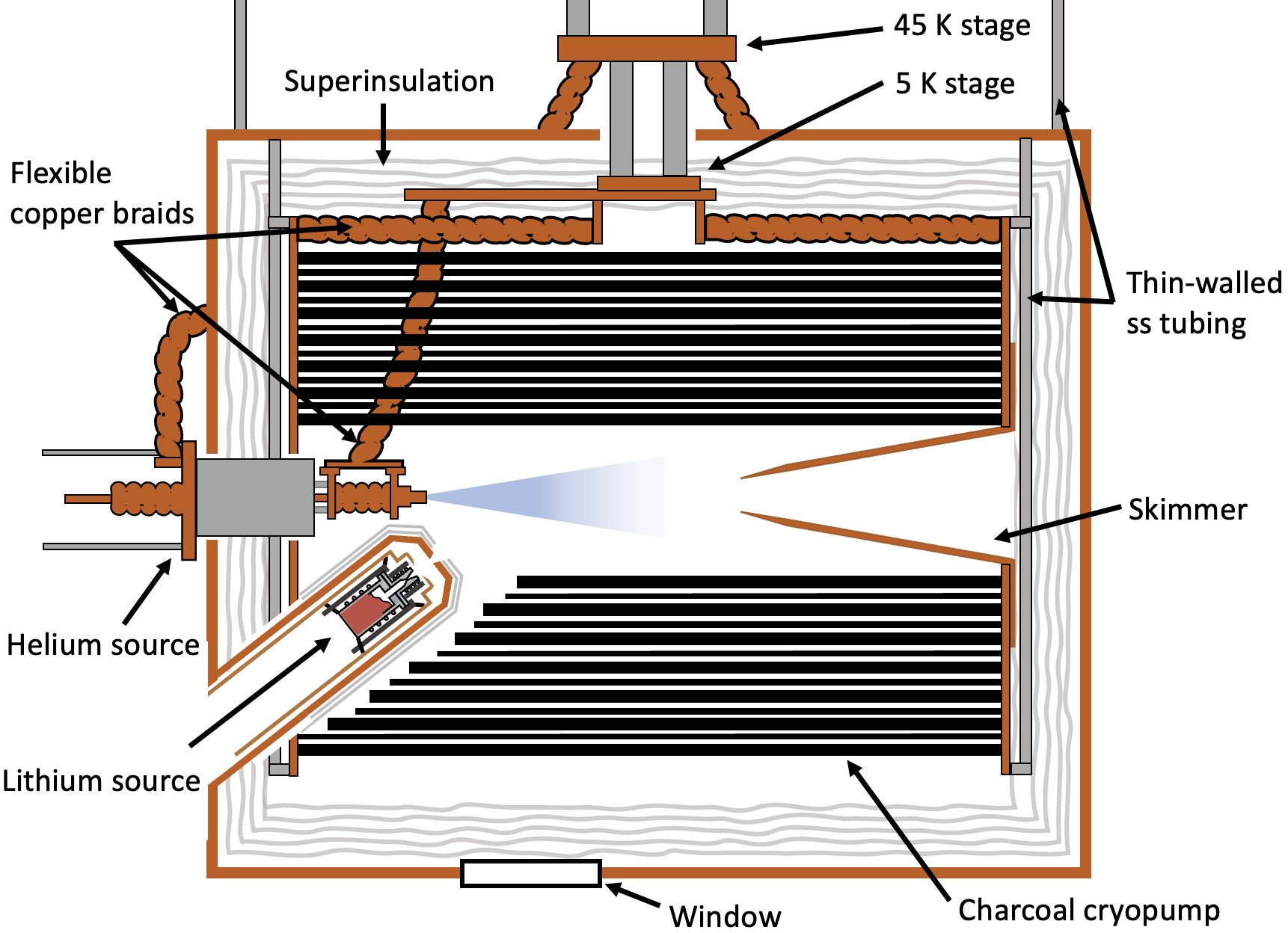}
    \caption{Schematic of the two stage cryogenic region}
    \label{fig:cryogenics}
\end{figure}

A schematic of the cryogenic region is shown in Fig. \ref{fig:cryogenics}. We cool our cryogenic components with a two-stage Cryomech model PT410 Pulse Tube refrigerator. Its first stage has a cooling capacity of 40 Watts at a temperature of 45 K, and its second stage has a cooling capacity of 1 Watt at a temperature of 4.2 K. The cryogenic region is surrounded by an 18x18x18 inch copper heat shield thermally connected to the refrigerator's first stage. This provides isolation of the cryogenic volume from room temperature blackbody radiation. The shield is suspended by thin walled stainless steel tubing from the vacuum chamber's ceiling to minimize thermal conduction.
Multiple layers of superinsulation are placed both outside (not shown) and inside this shield to reduce radiative heat loads \cite{insulation}.

Helium flows into the cryogenic region through a heat exchanger connected to the refrigerator's first stage, then through a second heat exchanger connected to the refrigerator's second stage, and finally through the nozzle where it expands into vacuum.

Thermal connections to the refrigerator cold plates are made by multiple flexible copper braids welded to clamps. Each braid is composed of 2880 36-gauge wires. The braids allow for spatial adjustments to the helium source during operation, and prevent the transfer of excessive strain to the PT refrigerator. 
To decrease thermal resistance, Apiezon N is applied at all thermal junctions. C10100 copper is used throughout to maintain conduction at cryogenic temperatures \cite{copperCryogenicResistivity}.

In order to avoid shock front formation and excessive helium background gas pressures, it is necessary to have a very high pumping capacity for the helium gas. In our experiment, this is provided by a charcoal cryopump. It consists of 60 rectangular 4x16 inch copper fins covered in an epoxied layer of charcoal. These fins are thermally connected to the second stage of the refrigerator, and typically operate at a temperature of 4.0 to 5.5 K. At these temperatures, the charcoal is a highly effective adsorption-based pump for helium gas \cite{Vasquez_1988_charcoal_pump}. The fin surfaces lie along lines pointing radially outward from the jet centerline. The rectangular openings between the fins form an approximately cylindrical pump opening that surrounds the expanding jet.  

Helium atoms have an adsorption energy on charcoal that varies with the quantity of adsorbed helium from 80 K to 300 K per atom \cite{Vasquez_1988_charcoal_pump}. This is the dominant heat load on the experiment, and sets the limit to our total helium flow rate. For instance if we budget 0.5 Watts for this heat load, the maximum flow rate with an adsorption energy of 200 K per atom is $1.8 \times 10^{20}$ atoms/s = 400 sccm.

About 1.2 to 1.4\% of the directed flow of the helium exits the cryogenic region through the skimmer, assuming the helium jet angular profile given in Ref. \cite{densityProfile}. Thus, the gas load on the following chamber is about 0.65 to 3.9 sccm (0.0082 to 0.049 Torr liters/sec) over our range of flow rates of 50 to 300 sccm. We pump this gas away with a CVC PVMS-1000 diffusion pump and a cooled chevron baffle that has an estimated net pump speed for helium of $S_2 = 3,500$ L/s.

To determine the pumping speed of the cryopump, we conducted an experiment wherein we blocked the line of sight between the helium nozzle and skimmer aperture. The gas load on the diffusion pump is then dominated by the flow of cryogenic helium background gas through the skimmer opening. The temperature of this gas must be intermediate between the 4.4 K and 45 K temperature of the cryogenic region surfaces; for purposes of this estimate we assume it is 15 K. Taking into account the measured pressure increase in the room temperature region and the known skimmer aperture conductance and diffusion pump speed, we estimate that the speed of our cryopump is  22,000 L/s, with an uncertainty of a factor of 2.

The estimated background gas pressure and density in the cryogenic region, with our maximum flow rate of 300 sccm, are $7 \times 10^{-6}$ Torr and $6 \times 10^{-18}  \: \text{m}^{-3}$. Using the He-He collision cross-section  \cite{Chrysos_2017_He_He_cross_section}  at T = 15 K  of $\sigma = 8 \times 10^{-15}  \: \text{m}^2$ , we estimate that the mean free path of a helium atom in the cryogenic region is about 20 cm.  With this pressure and mean free path, we expect our jet to transition from continuum to molecular flow without any formation of a shock front. We have seen no evidence of any effects of shocks in our experiments.

The charcoal has a finite capacity to adsorb helium. We find that we can flow helium continuously at a rate of 50 sccm for over $12$ hours. However, as the charcoal fills with helium the pressure in the cryogenic region begins to increase. We have observed that measured flux decreases with long run times and believe that this elevated pressure is the explanation. As the charcoal continues to fill with helium eventually a runaway condition occurs as elevated pressures promote thermal conduction which increases charcoal temperature. Eventually the charcoal climbs in temperature by tens of Kelvin, sheds its helium, and the run ends. The charcoal is then regenerated by pumping on it while it is warm. 

The gas load on our diffusion pump due to the directed flow of the helium is much larger than the flow of background helium gas through the skimmer. With a total flow rate of 300 sccm, the pressure above the diffusion pump rises to $P_2 = 1 \times 10^{-5} $ Torr. As discussed below, this room temperature background gas does limit the number of lithium atoms that we can extract from our source. We plan to eliminate this problem in a future upgrade of the apparatus.

\begin{figure}[t]
    \centering
    \includegraphics[width=\columnwidth]{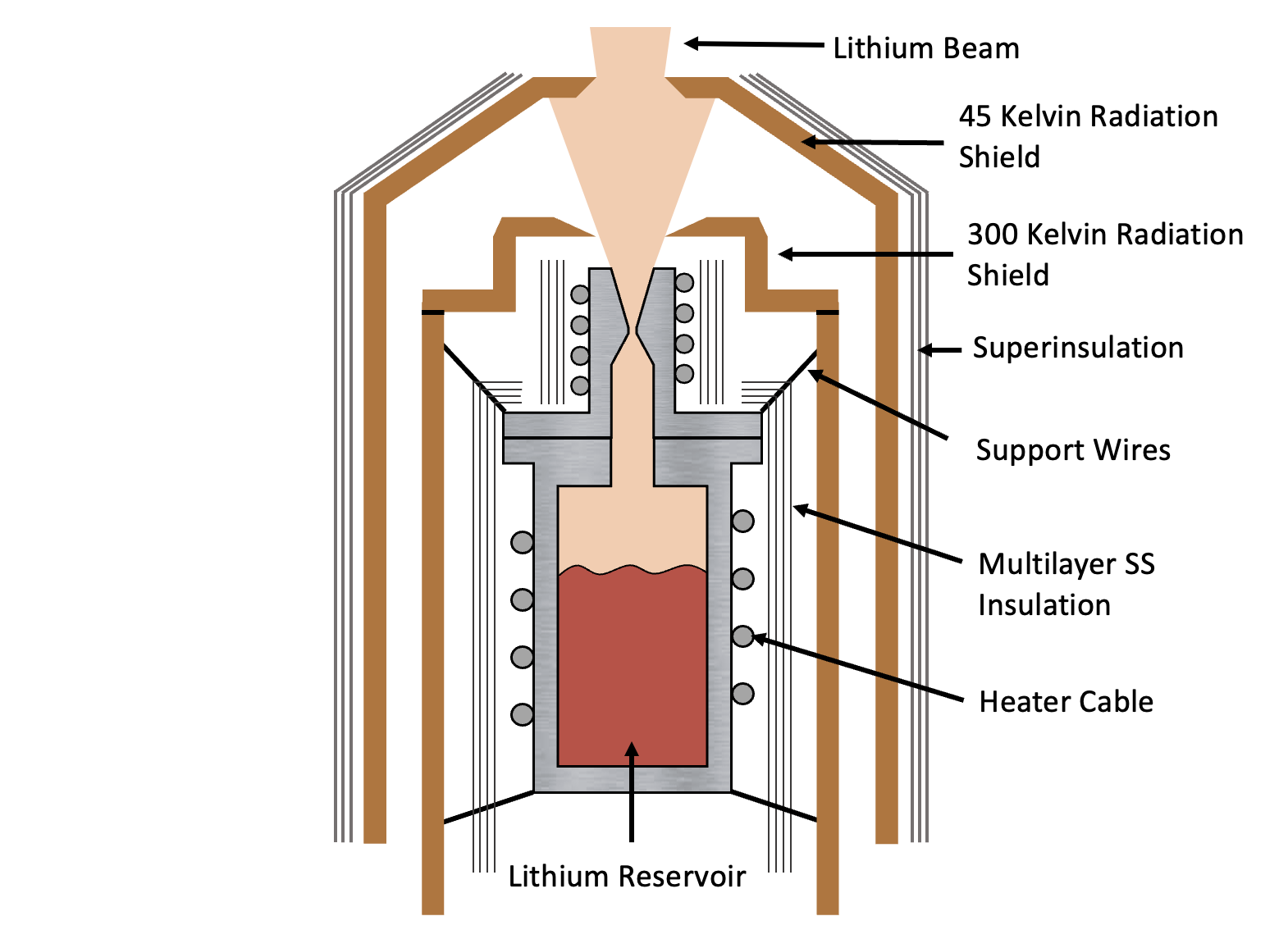}
    \caption{Schematic of lithium oven and insulation. The 1050 K oven is insulated from the cryogenics by several techniques.}
    \label{fig:oven}
\end{figure}

\subsection{Lithium source}

Our lithium oven is a two piece design with a separable reservoir and nozzle both heated independently as shown in Fig. \ref{fig:oven}.  Heating is generated by mineral insulated Inconel sheathed heater cables that are wound around and vacuum brazed to the oven. The oven is constructed from stainless steel (SS) and is suspended by SS wires inside a copper cylinder, which is cooled by room temperature water at the base. Multiple layers of thin SS sheets surround the oven providing a degree of radiation insulation. The copper cylinder resides inside a re-entrant extension of the 45 K outer structure. These insulating techniques allow the oven to operate at temperatures up to 1050 K with very minor heating of the cryogenics though they are separated by only 5.5 cm. The half angle of the extracted beam from the 45 K shielding is approximately 0.09 radians.

The two piece design allows us to fully disassemble the oven for refilling and to hold the nozzle at a higher temperature than the reservoir, which prevents nozzle clogging. The oven is loaded with $\sim$1.5 grams of lithium wire and can be run at its maximum temperature continuously for several hours. 

The oven nozzle is a conical design with a 1 mm diameter hole. When heated to $1050$ K, we estimate the Knudsen number to be $0.14$, in the intermediate region between supersonic and effusive. The oven was designed with a shaped nozzle to potentially generate a more forward directed flux, but the extent to which this is occurring is unknown.

\begin{figure}[t]
    \centering
    \includegraphics[width=\columnwidth]{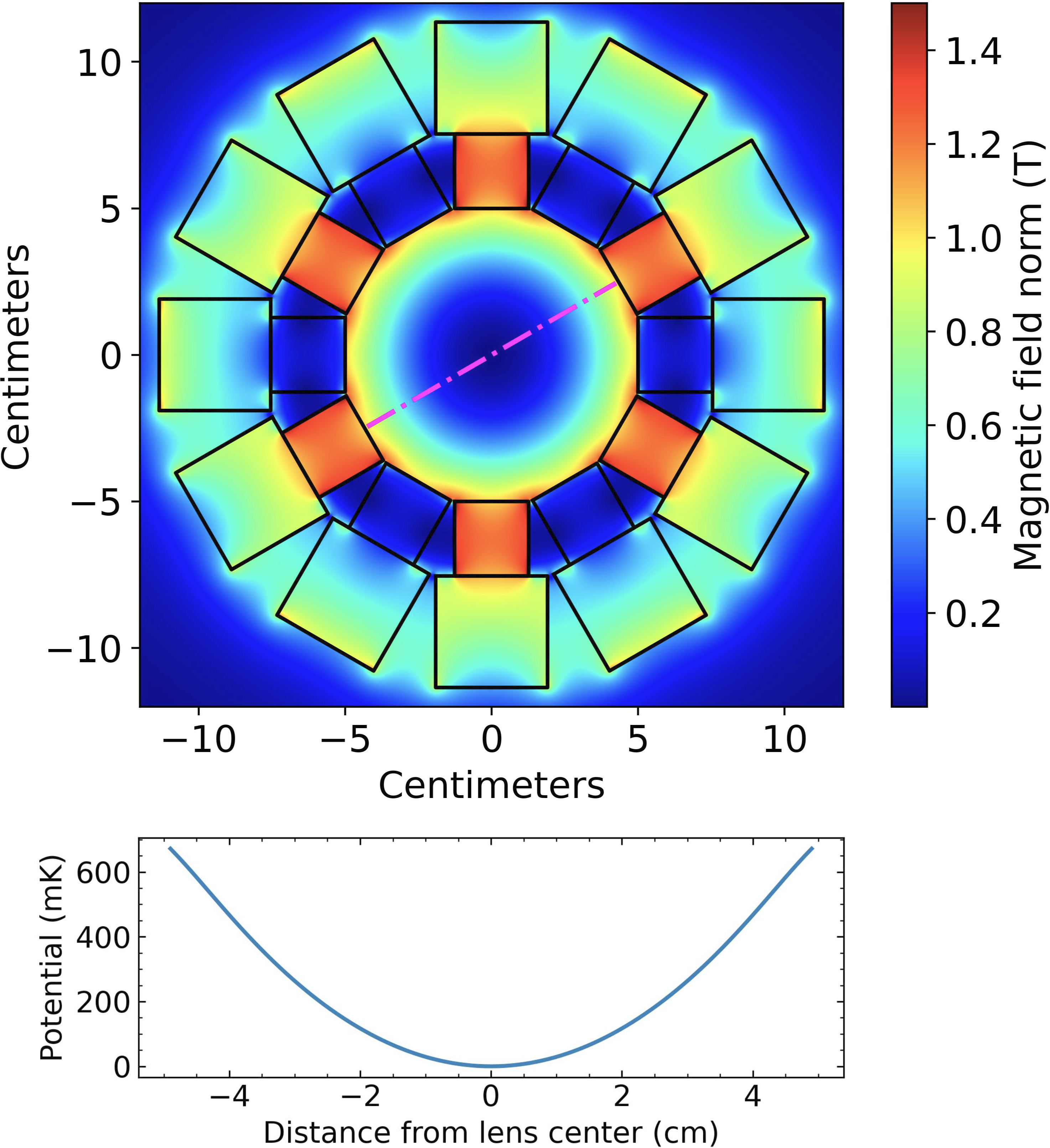}
    \caption{(a) Simulated norm of the magnetic field at the center-plane of our lens.  (b) A lithium atom's magnetic potential as a function of position along the dash-dotted line in (a).}
    \label{fig:lensField}
\end{figure}

\subsection{Atom lens}

The seeded lithium beam expands radially with the carrier gas and therefore falls in intensity. To maximise usable flux and intensity far from the nozzle, we capture and focus the lithium atoms with a magnetic hexapole lens \cite{Kaenders_1996_atom_lens, Chaustowsk_2007_magnetic_lens, Anciaux_2018_beam_brightening}. Helium, with no magnetic moment, continues expanding. 

A $2^2S_{1/2}\: \ket{F=2,m_F=2}$ state lithium atom in a magnetic hexapole field has potential energy given by

\begin{equation}
    V=-\vec{\mu} \cdot \vec{B}=\mu_B B_0\frac{r^2}{r_0^2}
\end{equation}

where $\vec{B}$ is the magnetic field vector, $\vec{\mu}$ is the magnetic moment vector, $\mu_B$ is the Bohr magneton, $r$ is the distance from the magnet axis, $r_0$ is a reference distance, and $B_0$ is $|\vec{B}|$ at $r_0$. This results in a harmonic restoring force over the length of the lens and is analogous to a gradient-index lens. 

Our atom lens contains a cylindrical Hallbach array of neodymium (NdFeB) permanent magnets \cite{Halbach_1980, Kaenders_1996_atom_lens}. The lens has a inner bore radius of 5.0 cm and length of 6 inches. It is composed of 6 slices, each 1 inch thick.  Each slice is made of an aluminum disk with an inner layer of twelve one inch cubic magnets and an outer layer of twelve 1.5x1.5x1 inch cuboid magnets epoxied into place, as illustrated in \ref{fig:lensField}(a). The magnets have a remanence of 1.26-1.29 Tesla. The coercivity of the inner layer magnets is 17 kOe, and that of the outer layer is 12 kOe, per the vendor's specification.

The atom lens's performance can be analyzed with ray tracing in a similar fashion to optical lenses. For a focusing metric, we chose the diameter of a circle which contains 90\% of the atomic flux, hereon out referred to as $D_{90}$. Using an original particle tracing code with simulated magnetic fields we traced an ensemble of $^7$Li atoms, all with velocity 210 m/s, originating from a point located $72$ cm from the lens face, which is the distance from the nozzle to the front facet of our magnet. Magnetic fields are calculated in 3D using analytic solutions \cite{magpylib}, and include demagnetizing effects. This calculation accounts for geometric aberrations only; in practice the lens focusing is also affected by the velocity distribution of the atoms as discussed below. 

\begin{figure}[t]
    \centering
    \includegraphics[width=\columnwidth]{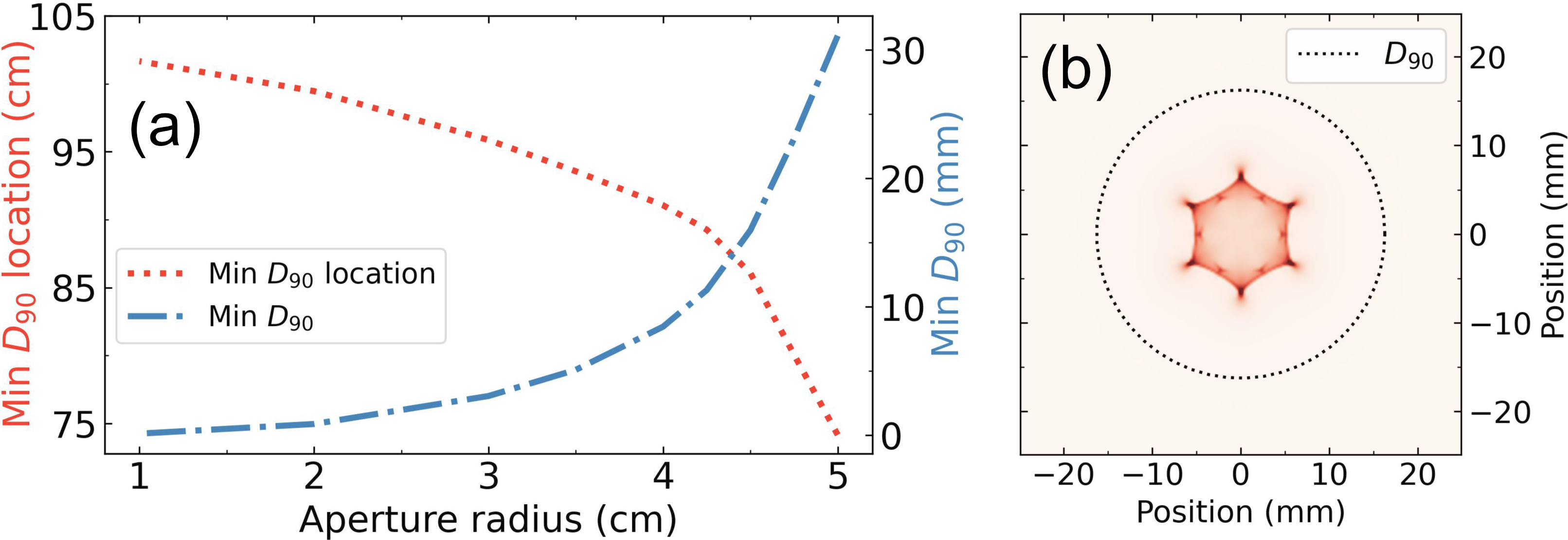}
    \caption{(a) Size and location of minimum $D_{90}$ versus aperture of atom lens for point source of particles. $D_{90}$ is diameter of circle containing 90\% of atoms, and location is distance from the end of the atom lens. (b) Heatmap of particle intensity at minimum $D_{90}$ for full lens aperture. Contrast was increased for readability.}
    \label{fig:lensFocus}
\end{figure}

Fig. \ref{fig:lensFocus}(a) shows the calculated position and size of the minimum $D90$ as a function of the radius of the lens open aperture. The results show a strong dependence on aperture radius. This is analogous to spherical aberration, and originates primarily from the fringing fields of the lens. To verify this, we also modeled the lens as a segment of an infinitely long lens, removing the effect of fringing fields, and found that the size of the minimum $D_{90}$ and the variation of the minimum $D_{90}$ location were substantially reduced.  

From this calculation, we also found that the maximum initial transverse velocity of atoms brought to the focus is 13.4 m/s. For our forward velocity of 210 m/s, the maximum half-angle for capture of the atoms by the lens is 0.064 rad. 

We have simulated the effect of collisions between Li atoms and He atoms in the directed flow inside the lens. This collision rate increases in the lens due to increased relative Li-He velocity. Our simulation shows that nearly all such collisions prevent the Li atom from reaching the focus. For this reason, the lens must be placed far enough from the nozzle that such collisions are improbable. The front face of our lens is positioned 72 cm from the nozzle, which is far enough that this condition is satisfied. A smaller  nozzle-to-lens distance could produce a somewhat higher focused flux due to a higher transverse capture velocity. However the magnet placement is still limited by collisions, and we do not expect that closer placement would result in a very large increase in focused flux.

\subsection{Optical Pumping}

\begin{figure}[t]
    
    \includegraphics[width=\columnwidth]{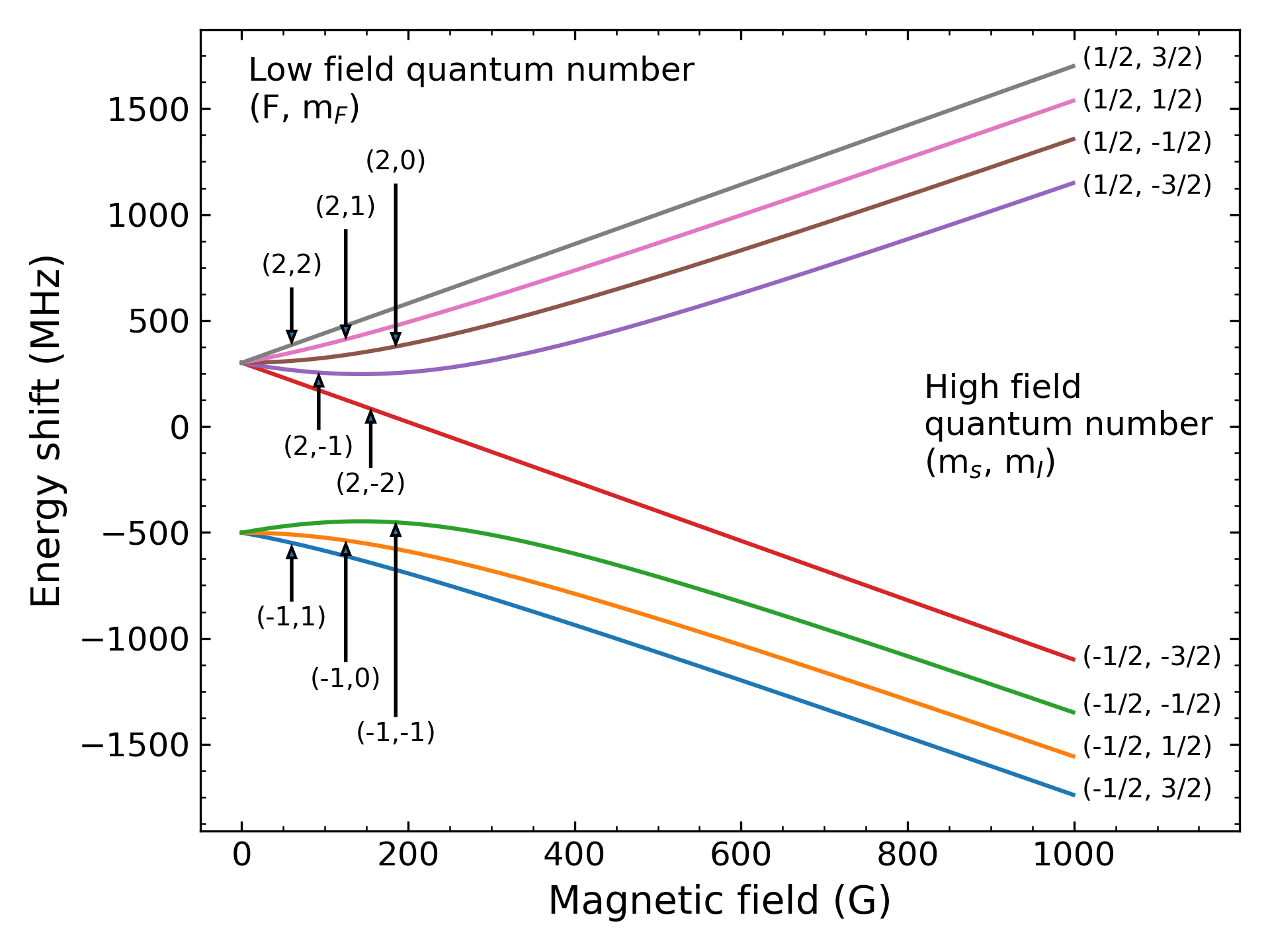}
    \caption{Breit-Rabi diagram for the $^7$Li $2S_{1/2}$ ground state levels. Depending on the atoms' position in the atom lens they can be in either the low field, high field, or intermediate region.}
    \label{fig:breit-rabi}
\end{figure}

The performance of the atom lens depends strongly on the state preparation of the atoms. Ideal focusing requires a linear restoring force. As shown in Fig. \ref{fig:breit-rabi}, half of the ground state hyperfine sub-levels are high field seeking states; these will be defocused and lost from the beam. Of the remaining states, only the $2^2S_{1/2}\: \ket{F=2,m_F=2}$ state experiences a perfect linear restoring force. The other three low-field seeking states will also be focused, but the focus position and size for these states differ from that of the $2^2S_{1/2}\: \ket{F=2,m_F=2}$ state. This introduces a form of aberration to the lens that will increase the size of the focus.

The lithium atoms initially have equal population in all eight states. In order to obtain maximal performance, both in terms of flux and focus size, we optically pump the atoms into the $2^2S_{1/2}\: \ket{F=2,m_F=2}$ state. Without optical pumping we find that the FWHM of the focus increases by approximately $30\%$ along with half the atoms being lost.

The maximum angular divergence of atoms that exit the skimmer have a Doppler shift of $\sim 20$ MHz. To optically pump all capturable atoms, the laser is frequency broadened. This is done by modulating the current supply of the laser at $800$ kHz with a modulation depth sufficient to widen the laser linewidth to $\sim 25$ MHz. Frequency broadening via current modulation has the additional benefit of reducing excess scattered photons as compared to power broadening. This is important at high densities as re-absorption of scattered photons could limit the effectiveness of hyper-fine state optical pumping.

\subsection{Laser-optical system}

The optically pumped atomic beam is characterized through fluorescence and absorption spectroscopy. For this we employ two lasers operating on the $2S \rightarrow 2P$ transition near $670.8$ nm, one for the optical pumping and the other for spectroscopy. Both are External Cavity Diode Lasers (ECDL) in a Littrow configuration. The ECDL used for optical pumping is part of a Master Oscillator Power Amplifier (MOPA) configuration with an Eagleyard Tapered amplifier and has a maximum output power of $0.5$ W. This laser is locked to the $(2^2S_{1/2}\:F=2 \rightarrow 2^2P_{1/2}\: F'=2)$ transition using a fluorescence signal from a reference atomic beam that is captured with a Photo-Multiplier Tube (PMT). The reference beam is generated in a separate chamber which contains an effusive oven. 

Since the laser used for spectroscopy is tuned (typically over ~4-5 GHz), we are unable to lock it to a fluorescence signal from the reference chamber. Instead the laser is locked to a temperature stabilized low finesse tunable etalon. To frequency calibrate the laser, we use the fluorescence signal from our reference atomic beam. The spectroscopy laser is sent angled into the chamber and retro-reflected. Scanning over the D2 line results in two equal but opposite Doppler shifts for both the $F=1$ and $F=2$ ground states. This produces dips between each pair of peaks. The separation between the dips is the ground state hyper-fine splitting which is used to calibrate the laser frequency to within a few MHz. 

For optical pumping, a repumper is required for atoms that transition to the $(2S_{1/2}\: F=1)$ ground state and would otherwise be lost. When it facilitates analysis by preventing unwanted optical pumping, a repumper is used with the spectroscopy beam. With both lasers, an electro-optic modulator (EOM) is driven at $803.504$ MHz, producing two side-bands one of which serves as the repumper. 

\section{Results}

\subsection{Near-field Atomic Beam}

The efficiency with which lithium can be seeded into the helium jet is characterized in the cryogenic region. We refer to this region as the ``near field" in contrast with the results at the atomic focus. Performing fluorescence spectroscopy in this region allows us to measure the velocity spread of the lithium atoms as well as the flux within the capturable solid angle of the magnet. These results can be compared with simulations and give crucial insight into understanding the conditions that maximize extracted flux.

\begin{figure}
    \centering
    \includegraphics[width=\columnwidth]{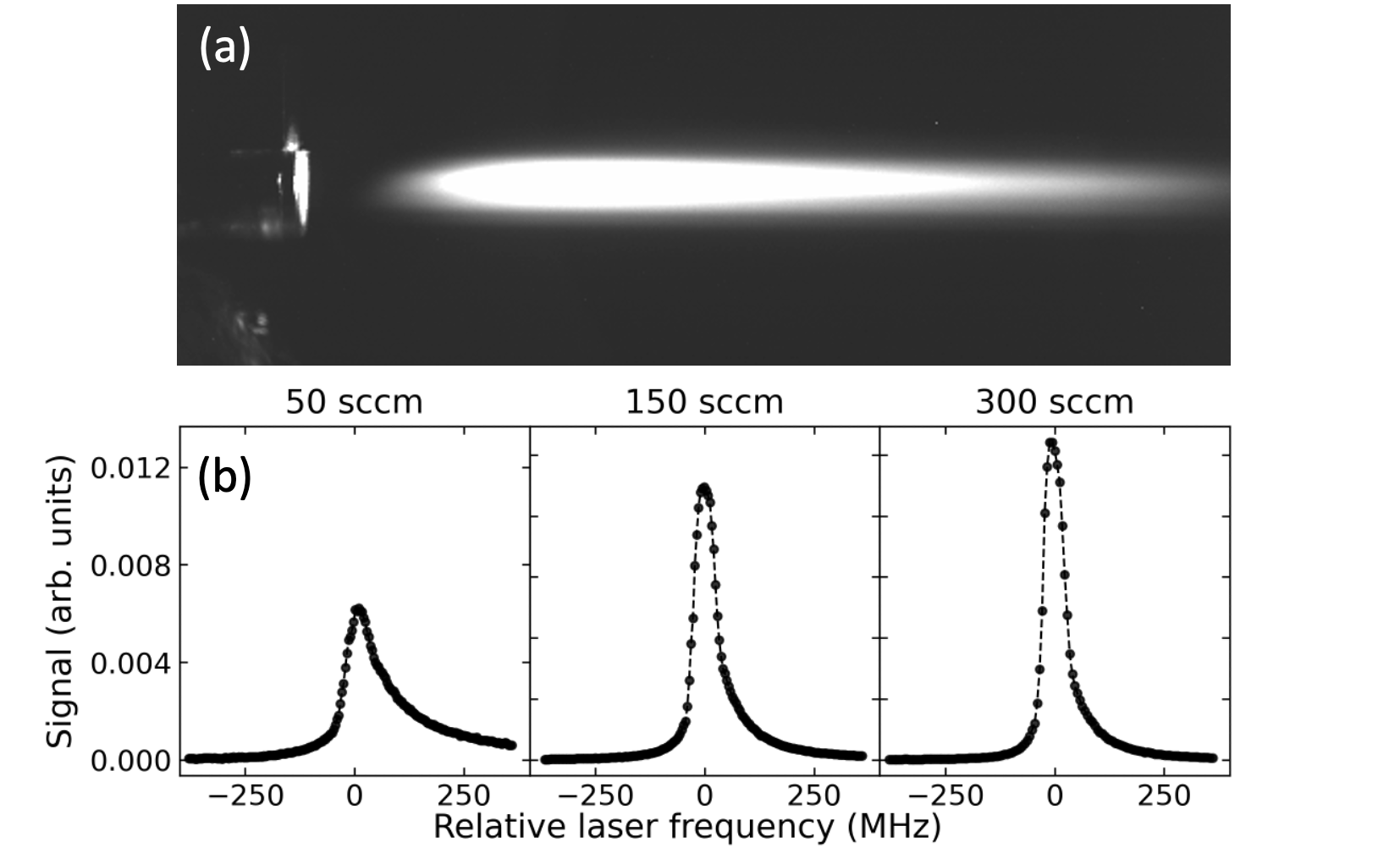}
    \caption{(a) Fluorescence image of the jet. Seeding position relative to the nozzle is approximately $1.7$ cm. Lithium atoms that have been entrained and translationally cooled within the jet are clearly visible. (b) Atomic fluorescence spectra results for various flow rates $4$ cm from the nozzle using a laser perpendicular to the atomic beam. The frequency scale is relative to the Doppler free signal of the D2 line from the reference chamber.}
    \label{flowRates}
\end{figure}

Fluorescence spectroscopy in the cryogenic region is performed by sending a laser down the center-line of the jet through the skimmer or vertically through a window in the cryogenic region. The signal is then imaged by a Ximea MC031MG-SY-UB CMOS camera. From the horizontal probe beam, the Doppler shift relative to a reference signal from the D2 line is used to determine the forward velocity, with the typical value being $210(2)$ m/s for flow rates below 250 sccm and $217(2)$ m/s for flow rates of 250 sccm of higher.

Spectral profiles for a range of helium flow rates at the centerline of the jet are given in Fig. \ref{flowRates}(b). The lineshapes are dominated by the Doppler effect, and therefore measure the vertical velocity distributions of the lithium atoms. A $100$ MHz Doppler shift corresponds to a velocity of $67$ m/s. The Doppler widths in Fig. \ref{flowRates} are a small fraction of the Doppler width of the lithium atoms from the oven. However, the lineshapes are also asymmetric, which indicates lithium atoms which still have a relatively high velocity compared to the expected velocity distribution of the jet.
Measurements of the longitudinal velocity distribution yield similar fluorescence profiles indicating lithium has not fully thermalized with the jet. If the lithium was fully thermalized with the helium at 1 mK, the observed FWHM would be comparable to the natural linewidth of 6 MHz. Within our viewing region the width of the spectral distribution decreases further from the nozzle. We expect that collisions continue to occur past 4 cm, the limit of our viewing region, reducing the asymmetry and width of the velocity distribution. 

\begin{table}[t]
\caption{\label{nearFieldFlux}Near field average $^7$Li atomic density within the capturable solid angle of the skimmer versus helium flow rate at an oven temperature of $800$ K with a seeding distance of $1.7$ cm from the nozzle. Data is collected 4.1 cm from the nozzle. The flux leaving the cryogenic shielding surrounding the lithium source was measured to be $\SI{1.4(2)e14}{}~\mathrm{s}^{-1}$ which allows for the simulated density to be computed.}

\begin{ruledtabular}
\begin{tabular}{l l l}
\thead{Flow Rate \\ (sccm)}&
\thead{Observed Density \\ ($\times10^{8}~\mathrm{cm}^{-3}$)}&
\thead{Simulated Density \\ ($\times10^{8}~\mathrm{cm}^{-3}$)}\\
\colrule
\\
50\footnote{Seeding distance of $0.7$ cm.}& 2.4(4) & 2.3(4)\\
50 & 1.1(2) & 1.0(2)\\
100 & 2.6(5) & 2.4(4)\\
150 & 4.5(8) & 3.2(5)\\
200 & 4.0(7) & 3.1(5)\\
250\footnote{\label{v_217}Terminal velocity of $217(2)$ m/s} & 3.7(7) & 2.6(4)\\
300\footref{v_217} & 3.1(6) & 2.1(3)\\

\end{tabular}
\end{ruledtabular}
\end{table}

Helium flow rate and seeding position, both of which are easily adjusted during operation, were varied to maximize near field density. Here we define the seeding position to be the distance from the nozzle to the centerline of the lithium beam, as shown in Fig. \ref{seedingExample}, which is determined by analyzing fluorescence images with zero helium flow. The maximum average density within the solid angle of the skimmer is obtained with a seeding distance of 1.7 cm with a helium flow rate of 150 sccm. The peak density $n_0$ is given by,

\begin{equation}
    \label{peakDensity}
    n_0 = \frac{\hbar \omega_0}{\sigma(w_0)} \frac{\Phi}{ \int \La(\vec{r}) I(\vec{r}) dV}
\end{equation}

\noindent where, $\La(\vec{r})$ is the spatial profile of the atomic beam normalized to a height of one, $I(\vec{r})$ is the laser intensity, $\sigma(w_0)$ is the resonant Doppler broadened cross section, and $\Phi$ is the number of scattered photons per second at resonance. The spatial profile is determined by fitting the fluorescence signal assuming cylindrical symmetry. 

Results for various flow rates are given in Table \ref{nearFieldFlux}. Due to the asymmetric distribution, the average density within the capture angle of the skimmer includes atoms which have not entirely thermalized with the jet. Since atoms with a high relative velocity to the jet are not focused by the hexapole magnet, it is not expected that the entirety of the atoms that escape through the skimmer will arrive at the atomic focus. Simulated results for the fraction of atoms that leave the skimmer and arrive at the focus are discussed in the modelling section and provided in Table. \ref{seedingEfficiency}.

\subsection{Focused Atomic Beam}

\begin{figure}[t]
    \centering

     \includegraphics[width=\columnwidth]{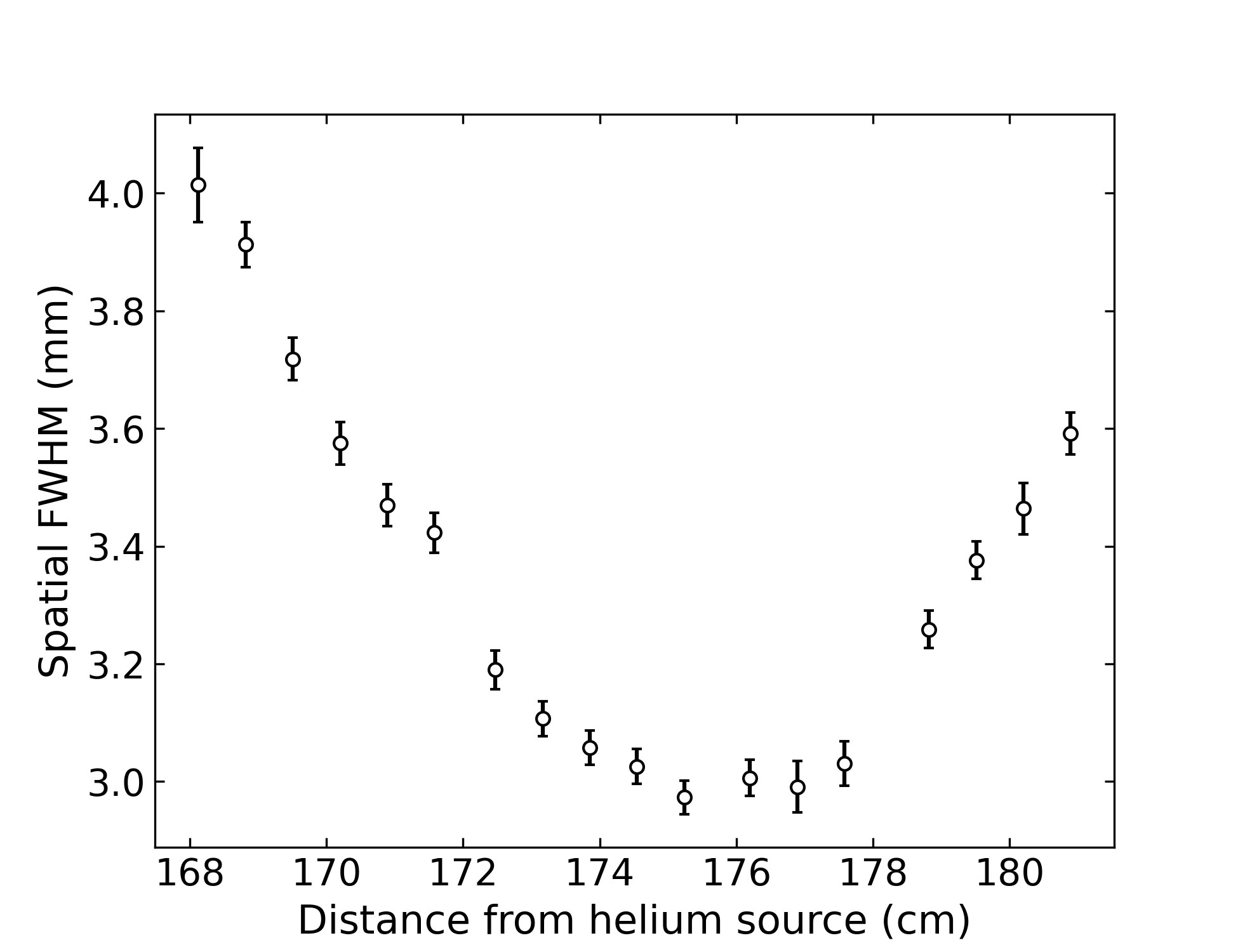}
 
    \caption{Spatial width of the atomic beam over a 14 cm distance in which a clear focus of the atomic beam is shown. The focus occurs approximately $176$ cm from the nozzle. The temperature of the lithium oven was $820$ K, and the beam has a forward velocity of 207(3) $\textrm{m}/\textrm{s}$.}
    \label{FWHM}
\end{figure}

At the focus, the atomic beam is characterized using both fluorescence and absorption spectroscopy. Absorption spectroscopy gives more reliable values for the atomic density than fluorescence spectroscopy, but is only appreciably present at lithium oven temperatures over 900 K. Fluorescence measurements are taken by a vertical laser beam passing through the atomic beam focus and viewed with a FLI ML1001 CCD camera viewing from the horizontal direction (see Fig. \ref{fig:Apparatus}). From this we can extract the spatial profile, transverse spectral profile, transverse temperature, flux and brilliance. Absorption measurements are taken with a horizontal beam passing through the atomic beam focus and imaged onto the CCD camera. 

For fluorescence spectroscopy with the vertical probe beam, the laser is tuned over the $\mathrm{F}=2$ transition on the $\mathrm{D}_2$ line of $^7\mathrm{Li}$. The beam is circularly polarized using a quarter wave-plate. A set of cylindrical lenses is used to produce a probe beam with a beam waist of $2.8(3)$ cm and $0.59(3)$ mm in the longitudinal and transverse directions respectively. This minimizes moving the vertical probe beam when characterising the focus as well as simplifying analysis. Similar to the optical pumping in the cryogenic region, sets of Helmholtz coils produce a $2$ G field aligned with the laser while reducing stray fields to $< 25$ mG. Since the atoms are optically pumped in the near field, the only allowed transition is $\ket{F=2,m_F=2} \rightarrow \ket{F=3,m_F=3}$. However, imperfect polarization or magnetic field alignment will lead to atoms transitioning to other states. To limit these effects, the laser intensity is kept sufficiently low (typically $< 1~\mathrm{\mu W}/\mathrm{cm}^2$) such that atoms scatter no more than one photon on average. 

\begin{table}[b]
\caption{\label{focusedResults}%
Focused beam characteristics with fluorescence and absorption imaging at an oven temperature of $1030$ K. For brightness and brilliance the FWHM of the velocity distributions are used to define the FWHM of the angular distributions. These quantities, along with the density and intensity, correspond to the peak spatial value. All measurements are taken at the atomic focus with the exception of the longitudinal velocity spread which is measured approximately $5$ cm from the focus. 
}

\begin{ruledtabular}
\begin{tabular}{lcc}
\textrm{Quantity}&
\textrm{Fluorescence}&
\textrm{Absorption}\\
\colrule
\\

FWHM (mm) & 5.01(4) & 4.41(3)\\
Density $(\textrm{cm}^{-3})$ & \SI{3.4(6)e8}{} &\SI{2.64(9)e8}{}\\
Intensity $(\textrm{cm}^{-2}\textrm{s}^{-1})$ & \SI{7(1)e12}{} & \SI{5.6(2)e12}{}\\
Flux\footnote{Flux within a $1$ cm diameter circle.} (s$^{-1}$) & \SI{2.3(4)e12}{} & \SI{1.54(6)e12}{}\\
Brightness $(\textrm{m}^{-2}\textrm{s}^{-1}\textrm{sr}^{-1}$) & \SI{2.4(8)e19}{}\\
Brilliance $(\textrm{m}^{-2}\textrm{s}^{-1}\textrm{sr}^{-1}$) & \SI{1.6(5)e21}{}\\
$T_{\perp}$ (mK) & $<$ 20\\
$T_{\parallel}$ (mK) & 7(3)\\
$v_{\mathrm{terminal}}$ (m/s) & 211(2)

\end{tabular}
\end{ruledtabular}
\end{table}

At high lithium oven temperatures the extracted flux is sufficiently dense to allow for absorption spectroscopy. A probe laser with $803.504$ MHz side-bands is imaged directly on the CCD and images are taken with and without the atoms. In order to ``switch" off the atoms, the helium flow is turned off. This process takes about 10 seconds for the flow to completely cease.

Results for fluorescence and absorption spectroscopy are given in Table \ref{focusedResults}. Fluorescence data is analysed in a similar manner to near field results using Eq. \ref{peakDensity} modified for circularly polarized light. The peak density from absorption data is determined from

\begin{equation}
    \label{absorptionEq}
    n_0  = \frac{- \ln(T)}{\sigma(\omega_0)\int \La(x) dx} 
\end{equation}

\noindent where $\La(x)$ is the spatial profile along a line through the center of the atomic beam normalized to a height of one and $T$ is the minimum transmittance. The spatial profile is determined from the inverse 
Abel transform of the absorption image.
Due to weak absorption off resonance we determine $\sigma(w)$ from fluorescence results. The flux is given as the atoms per second passing through a $1$ cm diameter circle.

\begin{figure}[t]
    \centering

     \includegraphics[width=\columnwidth]{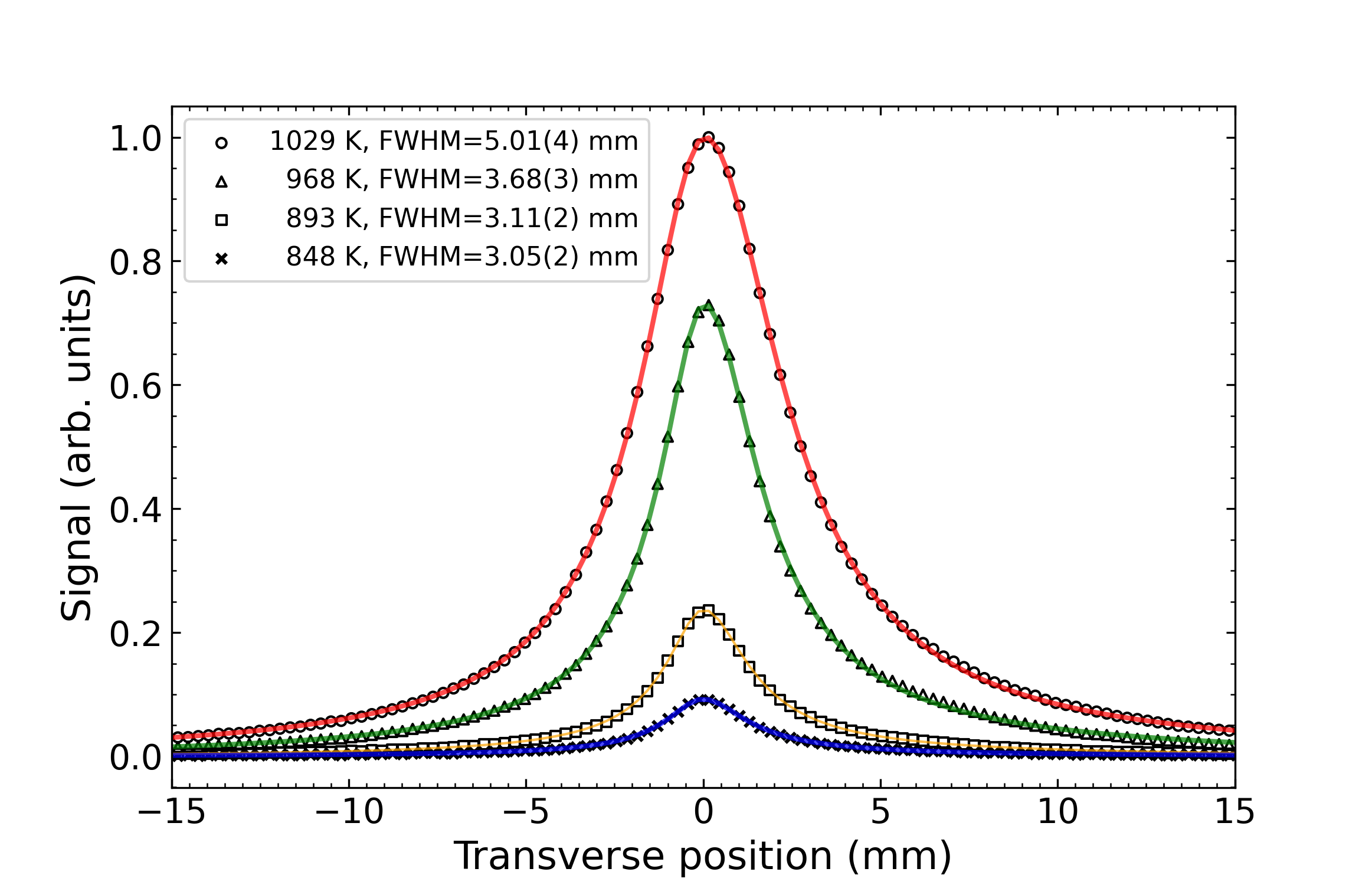}
 
    \caption{Spatial profile at the atomic focus for different oven temperatures fit to a q-Gaussian distribution. The data is from fluorescence spectroscopy and is normalized to the peak signal at the highest oven temperature.}
    \label{spatialProfile}
\end{figure}

\begin{figure*}[t]
    \centering
     \includegraphics[width=2\columnwidth]{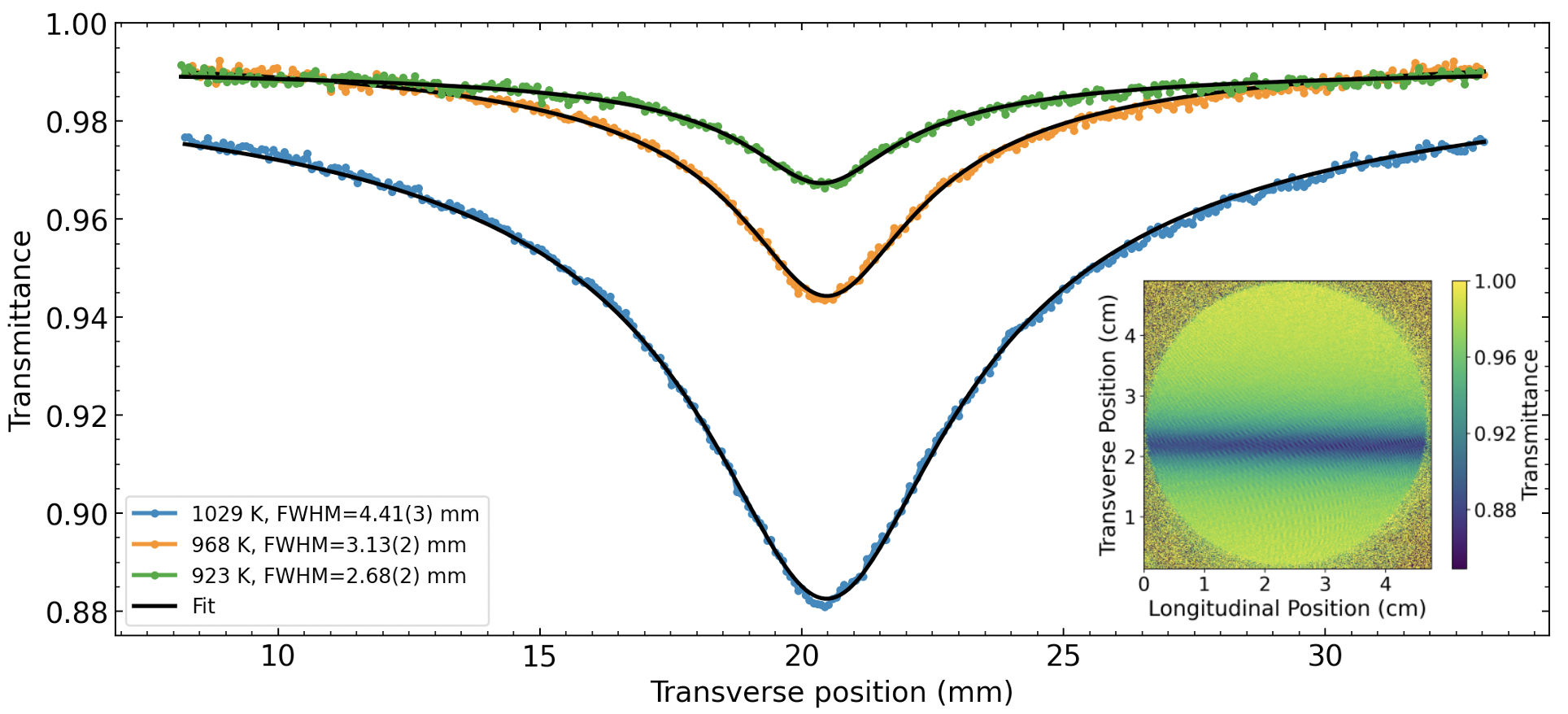}
    \caption{Absorption spectroscopy results at the atomic focus. The data is fit to the Abel transform of a Lorentzian profile to determine the FWHM of the atomic beam. In the bottom right corner is an example absorption image.}
    \label{absorptionProfile}
\end{figure*}

\begin{figure}[t]
    \centering
     \includegraphics[width=\columnwidth]{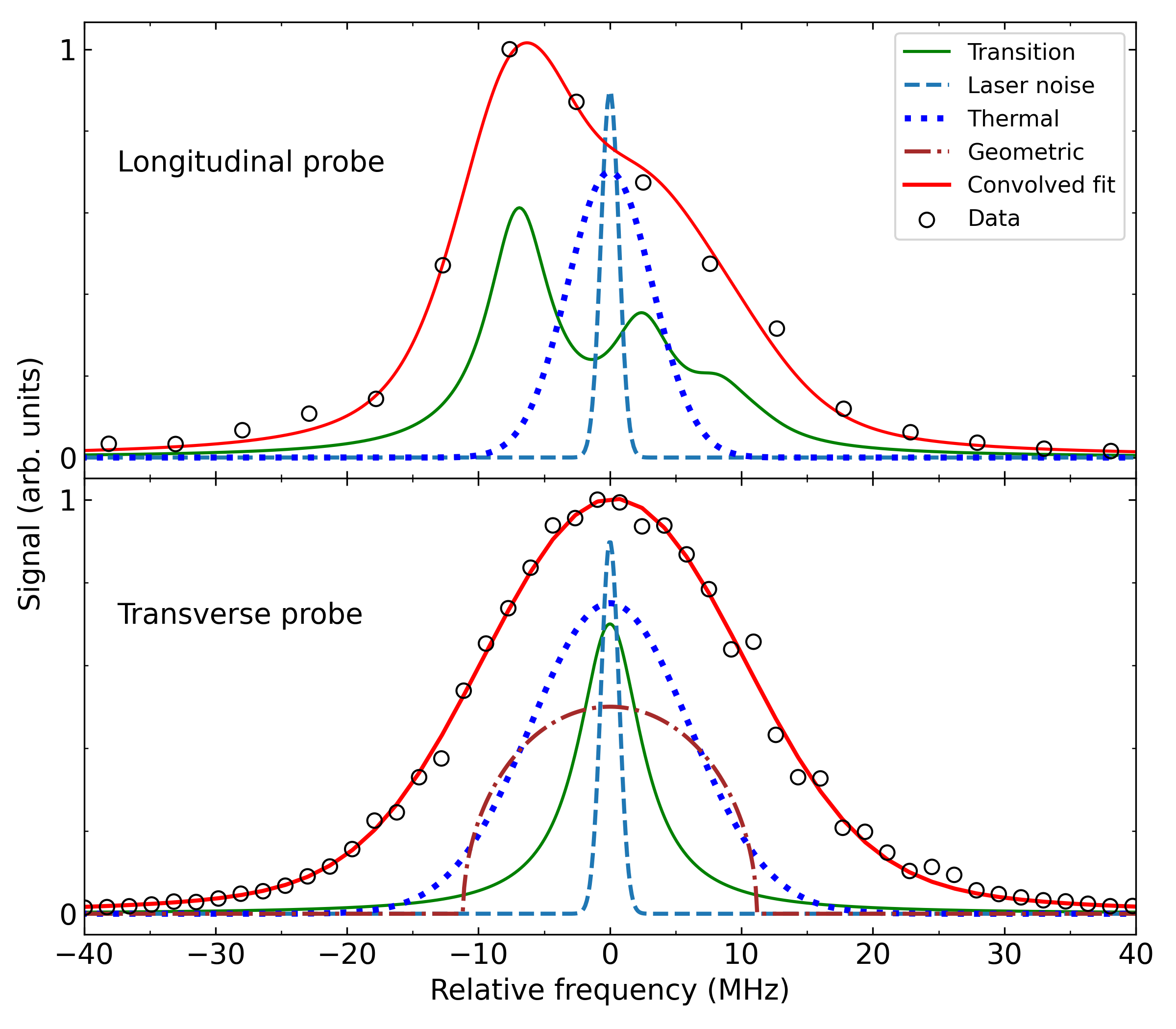}
    \caption{Data, fits, and underlying convolving profiles for fluorescence data at the  location of atom lens focus for (a) longitudinal probe laser and (b) transverse probe laser. Heights are arbitrary. The $2^2S_{1/2} \rightarrow 2^2P_{1/2}$ transition results in 6 hyperfine transitions yielding an asymmetric spectral profile as seen in (a). Optical pumping with a circularly polarized probe beam yields a single transition profile in (b). The geometric profile in the transverse direction models the output convergence angle of the atom lens. From the Doppler profile we can extract the temperature of the atoms. }
    \label{fig:farFieldProfile}
\end{figure}

To find the atomic focus, the laser and camera are mounted on a movable platform that can be adjusted during the experiment. The platform and viewing region allow us to collect fluorescence data over a 14 cm range. From particle tracing simulations it was expected that the size of the focus would remain relatively constant over a few centimeter region. We take the location of the minimum spatial FWHM to be the location of the focus. The spatial width as a function of distance from the atom lens is given by Fig. \ref{FWHM}. At each position the fluorescence data is fit to a Tsallis q-Gaussian distribution which we found fits the spatial data well \cite{Tsallis_1988, Douglas_2006_Tsallis_distribution}. A q-Gaussian allows for fitting to Gaussian like distributions, but with variable sized tails.

At the focus, fluorescence data is collected at various oven temperatures, as presented in Fig. \ref{spatialProfile}. As can be seen, the signal and the spatial width increase with oven temperature as was also found with absorption results given in Fig. \ref{absorptionProfile}. We have found that the near field spatial profile also increases in width with temperature suggesting that this effect is related to the seeding dynamics. Presumably, the additional heat load to the jet from the incoming lithium atoms is the culprit, although the exact dynamics are not well understood. While the size of the focus increases with lithium source temperature we have found that the location of the focus remains invariant to lithium source temperature. 

The longitudinal velocity profile and temperature is measured by a probe beam at a small angle to the centerline of the beam of 13 degrees (see Fig. \ref{fig:Apparatus}). This is to avoid the optical pumping and slowing effects possible when sending a laser down the entire length of the atomic beam. Because the probe laser is tilted, the transverse profile can broaden the longitudinal profile, particularly if measured at the focus. To minimize this, longitudinal measurements were taken 5 cm past the focus. A repumper laser is present yielding 6 total hyperfine transitions, 3 for each of the 2 ground states. The spectral profile is fit to a convolution of a Gaussian profile representing the thermal distribution, a Gaussian profile for the measured laser jitter, and 6 Lorentzian profiles at each hyperfine transition frequency with proper weighting. From the Gaussian profile the temperature can be extracted, which we find to be 7(3) mK. Results and convolving profiles are shown in Figure \ref{fig:farFieldProfile}(a).

The measured longitudinal velocity distribution at $z = 4 \: \mathrm{cm}$ in the near field does not show a single, sharp peak, but rather shows an asymmetric lineshape similar to the ones shown in Fig. \ref{flowRates}; the velocity spread at this point corresponds to a temperature of several tens of mK. The lower temperature of the extracted beam results from additional cooling by helium collisions for distances $z > 4 \: \mathrm{cm}$, and may also have a contribution from a velocity filtering effect by the magnetic lens.

We measured the transverse Doppler lineshape of atoms that transit through a 1 cm diameter circle centered on the beam focus using the fluorescence induced by the vertical laser beam. The measured lineshape is shown in Fig. \ref{fig:farFieldProfile}(b).
From this data, we determine that the FWHM of the transverse velocity distribution is 13.4(1) m/s. This implies that the beam emerging from the focus has an angular width of 0.064(1) rad (FWHM). 

The angular divergence of the beam, or equivalently its transverse velocity distribution, arises from two sources. One is the angular spread of the atomic trajectories arriving from the lens aperture, and the other is the random spread in transverse velocities arising from non-zero transverse temperature. Thus, the transverse Doppler lineshape is a convolution of four functions: the natural lineshape, the laser lineshape, the geometric Doppler lineshape that results from the distribution of ray paths from the lens aperture, and a thermal transverse Doppler distribution. 

The geometrical distribution of transverse velocity, not accounting for lens aberrations, is 

\begin{equation}\label{geometricEffectEq}
\begin{split}
    n(v_\perp) \approx \sqrt{1-(v_\perp/v_{\perp 0})^2 }
\end{split}
\end{equation}

where $v_\perp$ is the transverse velocity and $v_{\perp 0}$ is the maximum transverse velocity exiting the atom lens. Due to the substantial geometric aberration in our lens, atoms transiting the lens at larger radii do not contribute significantly to the flux within the 1 cm diameter circle. For this reason, we adopt a simplified model in which only atoms up to some maximum radius in the lens contribute to the signal, and correspondingly take $v_{\perp 0}$ as a fit parameter in this model. 

The best fit result for the overall lineshape is shown in \ref{fig:farFieldProfile}(b), along with the four contributing lineshapes. From this fit, we extract a transverse temperature of 14 mK. Considering possible variations of our model, we estimate an upper bound to the transverse temperature of 20 mK.

\section{Modeling}

\subsection{Monte-Carlo Simulation of Atomic Capture}

To determine the parameters that maximize seeding efficiency, a 3D Monte-Carlo simulation was performed. In the simulation, the helium jet is treated as a directed flow with a density profile based on early results of supersonic free jets by Ashkenas and Sherman and later verified by Tejeda, Fernández-Sánchez, and Montero \cite{densityProfile, shockFronts}. Lithium atoms are generated at the oven with velocities sampled from an effusive source and angles limited to the those which pass through the 45 K radiation shielding, as shown in Fig \ref{fig:oven}. The lithium atoms are then propagated through the jet with time steps that are small compared to the local inverse collision rate. At each time step, a collision is determined based on an acceptance rejection method using the local density, collision cross-section, and relative velocity. Collisions are evaluated in the center of mass frame with a scattering angle sampled from the distribution of the differential scattering cross section at the relative collision energy \cite{Makrides_2020_CrossSection, Tiesinga_2021_CrossSection}. As small scattering angles are most probable at high relative collision energies, it is possible for particles to undergo many collisions without becoming thermalized. This is shown in Fig. \ref{seedingExample} with a particle passing through the jet and undergoing collisions without experiencing a significant deflection in its trajectory.  For the simulation we neglect $^7$Li-$^7$Li interactions as these are expected to occur at negligible rates as well as the formation of shock fronts. Additionally, we assume that the lithium atoms act as a small perturbation to the jet, neither heating the jet appreciably nor altering its density profile.

\begin{figure}
    \centering

    \includegraphics[width= \columnwidth]{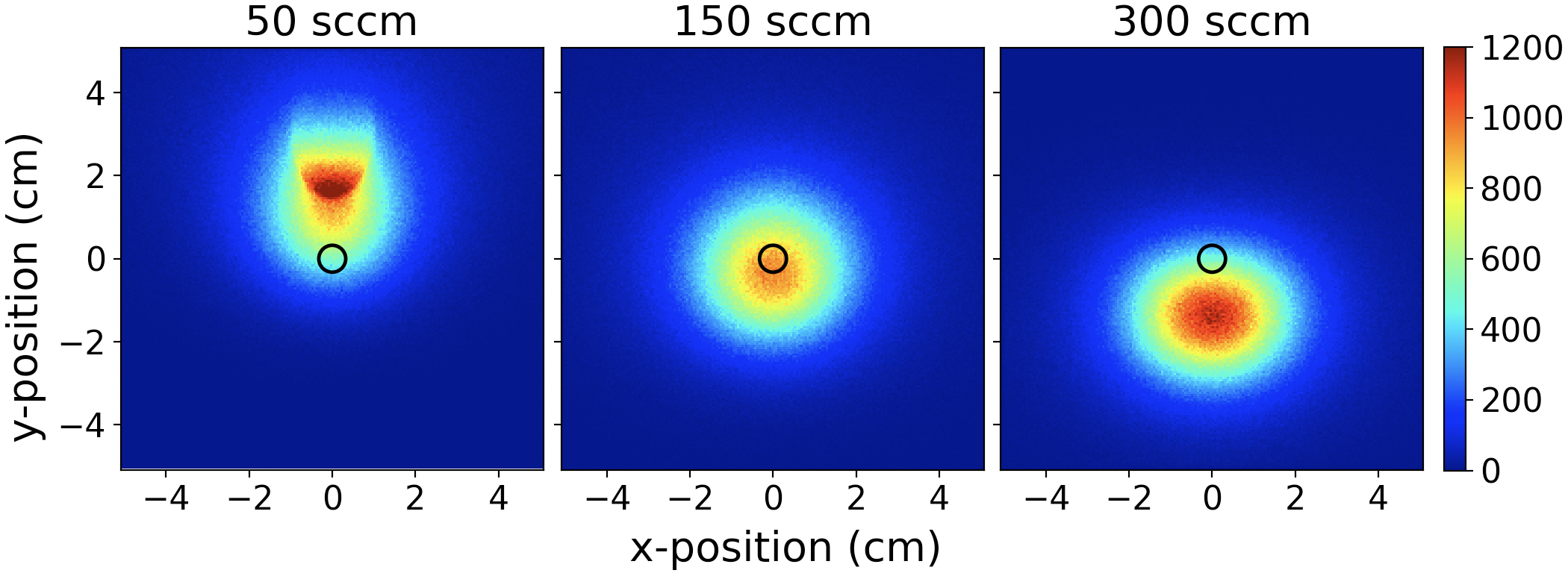}

    \caption{2D histogram of $5\times 10^6$ simulated particles and the positions of all particles that pass the xy plane $4.1$ cm from the nozzle. The seeding distance is $1.7$ cm and the black circle indicates particles that are within the capturable solid angle of the skimmer projected to the xy plane. Data is binned into $0.25 \times 0.25$ mm boxes. The number above each plot gives the helium flow rate.}
    \label{sim_flowRates}  
    
\end{figure}

Two parameters that can be adjusted experimentally are the helium flow rate and seeding distance relative to the nozzle, as shown in Fig. \ref{seedingExample}. These parameters were varied in the simulation, and using the measured flux leaving the lithium oven shielding, the average density within the capture angle of the skimmer is determined to allow for comparison with experimental results. The maximum average density occurs approximately $1.7$ cm from the nozzle with a flow rate of $150$ sccm, matching experimental results. Results for various flow rates at this seeding distance are given in Fig. \ref{sim_flowRates}. The effects of the collisional thickness of the jet are present with atoms penetrating through at low flow rates while being deflected downwards at high flow rates. At the lowest flow  rate of 50 sccm the shadow of the lithium source can be seen.

Due to effects from helium background gas, to maximize focused lithium we must operate at lower flow rates as discussed below. To compensate, we can seed closer to the nozzle with a seeding distance of approximately $0.7$ cm. As shown in Table \ref{nearFieldFlux}, this increases the density relative to the 1.7 cm distance, but doesn't provide the density we'd get if we were able to operate at higher flow rates while seeding further from the nozzle.

\begin{table}[b]
\caption{\label{seedingEfficiency} Simulated results for the percentage of atoms that arrive at the atomic focus from the lithium source.}
\begin{ruledtabular}
\begin{tabular}{l l l l l l l l}
Flow Rate (sccm) & 50\footnote{Seeding distance of 0.7 cm. All other results are for 1.7cm.} & 50& 100& 150 &200 & 250 &300\\
Efficiency (\%) & 0.11 & 0.04 & 0.21 & 0.37 & 0.43 & 0.25 & 0.20

\end{tabular}
\end{ruledtabular}
\end{table}

\subsection{Modeling the focus}

Our jet-seeding simulation can be combined with a simulation that traces particle trajectories through the atom lens to predict the shape and location of our atomic beam focus. To do this, we extend the Monte-Carlo simulation of the capture to a distance of 10 cm from the nozzle as further $^4$He-$^7$Li collisions are expected to occur past where our optical access ends. We assume that the helium gas continues to cool adiabatically out to this distance. Beyond this distance and until the magnet, we estimate that there is less than one collision on average per lithium atom for helium flow rates of 50 sccm. We take the positions and velocities of the lithium atoms at a 10 cm distance as an input condition for calculating the trajectories through the lens. 

\begin{figure}[t]
    \centering
     \includegraphics[width=\columnwidth]{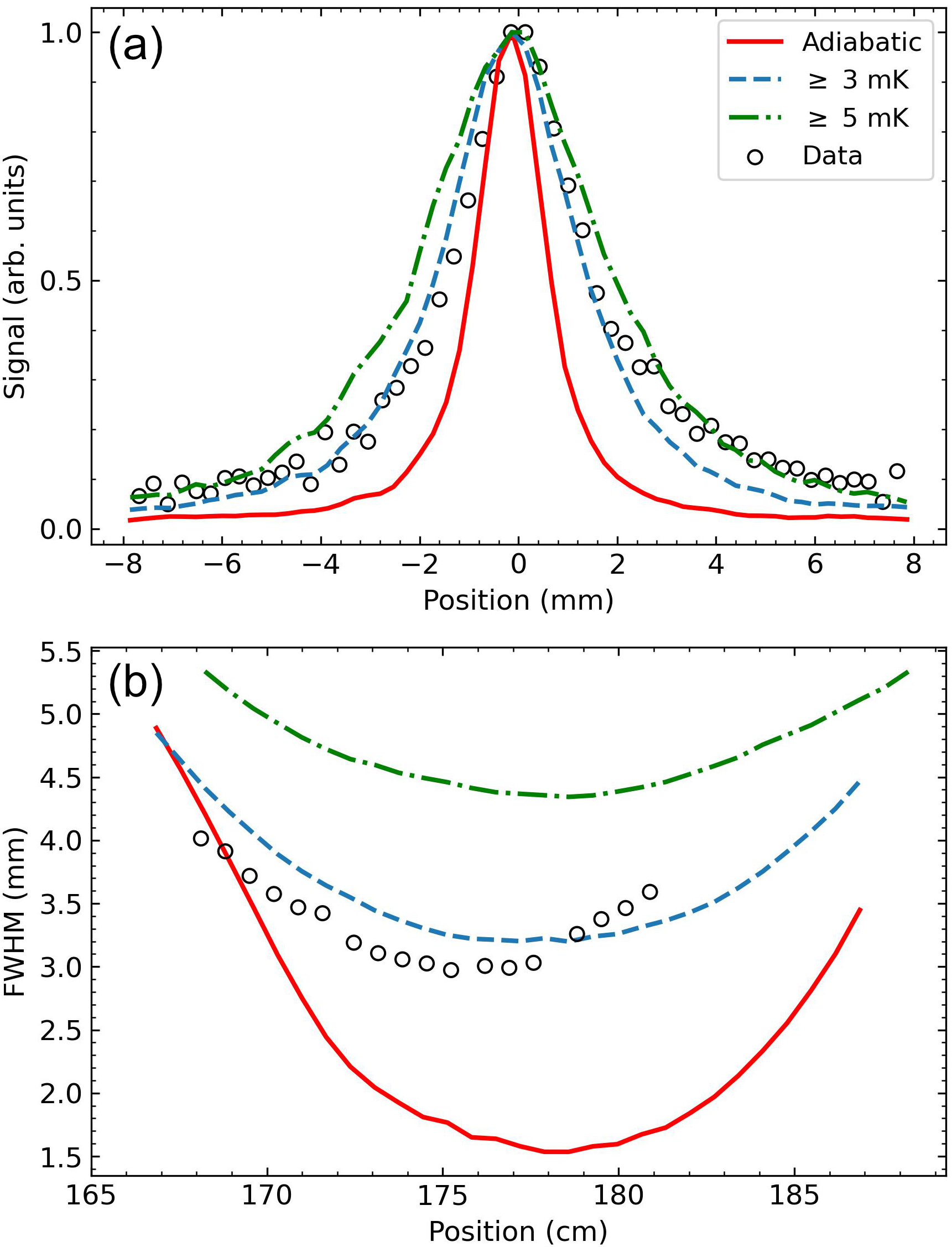}
    \caption{Simulated focus results versus flourescence data. (a) Measured transverse profile vs simulated transverse profile. (b) FWHM versus distance from nozzle for data and simulation. In the legend adiabatic refers to a pure adiabatic expansion, while 3 and 5 mK refer to introducing a temperature floor in the expansion.}
    \label{fig:focusVsData}
\end{figure}

Though our simulation does not account for heating or other phenomena which may affect the jet temperature, we can attempt to roughly model temperature effects by limiting the minimum temperature to which the jet cools. This causes a significant increase in the size of the focus for even a few millikelvin over the expected temperature, as shown in Fig. \ref{fig:focusVsData}. 

This effect is explained by a virtual source \cite{quittingSurfaceBeijerinck}, an imaginary source of atoms located at the nozzle plane that would produce the same spatial and velocity distribution at a given distance from the nozzle neglecting collisions. The phase space distribution of the virtual source is produced by projecting atoms back to the plane of the nozzle. Higher temperatures produce larger velocity distributions which in turn result in a larger virtual source size. With the virtual source as the object, our lens then produces a larger image for higher jet temperatures. 

Better agreement is found with elevated temperatures, but a physical model is lacking to motivate a specific temperature and thus profile. However, the fact that relatively minor and plausible temperature changes of the jet causes fairly significant changes in the focus size hints that additional dynamics may explain the discrepancy between simulation and experiment.

In addition to the location of the focus and FWHM at the focus, the simulations are used to predict the seeding efficiency or fraction of atoms leaving the oven shielding that arrive within a 1 cm diameter at the focus. These results are presented in Table \ref{seedingEfficiency}. Accurate measurements of the flux leaving the oven shielding is performed at lower temperatures to avoid absorption effects. At an oven temperature of 800 K, a flux of $\SI{1.4(2)e14}{}~\mathrm{s}^{-1}$ is measured leaving the oven shielding while a flux of $\SI{3.8(7)e10}{}~\mathrm{s}^{-1}$ is measured at the focus. This gives a total efficiency of 0.03\% and accounting for a 2.5 times loss from background pressure, as discussed below, agrees within 40\% of the simulated results.

\subsection{Effects of background gas pressure}

A major design concern is sufficient vacuum pumping to reduce scattering between lithium atoms and background gas helium atoms to negligible levels. We have determined that this condition is not satisfied. This was made clear, for instance, by the fact that our focused beam reached its maximum intensity at helium flow rates of about 50 sccm, while the near field flux was maximum for much higher helium flow rates of about 150 sccm. This apparent discrepancy is explained by increased background gas pressures and collisional loss rates of lithium atoms at the higher helium flow rates.

\begin{figure}[t]
    \centering
    \includegraphics[width=\columnwidth]{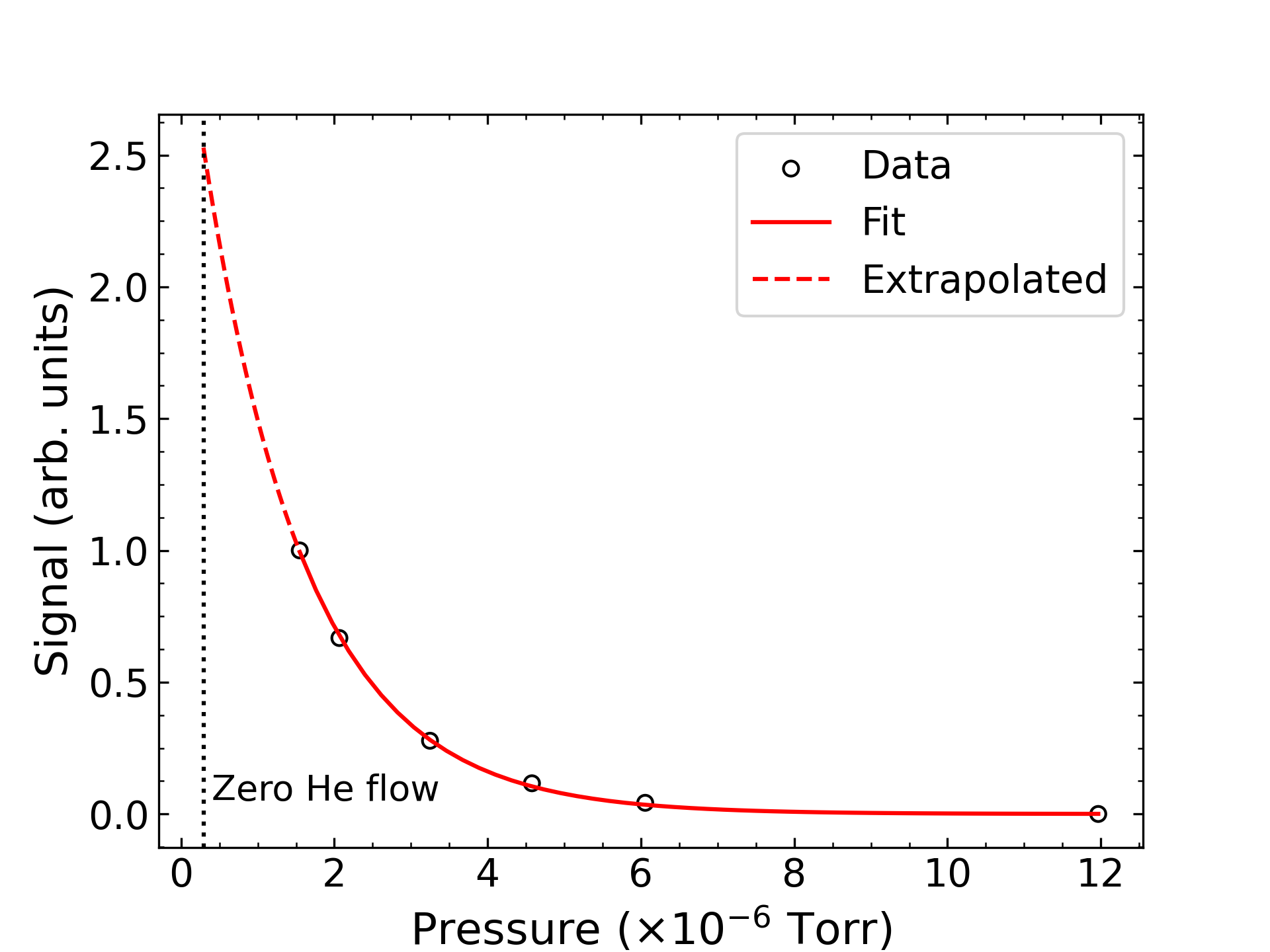}
    \caption{Peak signal at the atomic focus normalized to the maximum observed signal vs. background chamber pressure. The data was collected with $50$ sccm of flow from the nozzle. To increase the pressure, room temperature helium is bled into the chamber. The results show that the effects of collisions with background helium atoms are substantially reducing the flux at the atomic focus. Fitting the data to Beer's law and extrapolating the results to zero pressure indicates a $\sim 2.5$ times increase in flux. Note that the lowest pressure value is the pressure only from the supersonic nozzle flow.}
    \label{fig:vacuum}
\end{figure}

In order to determine the extent of this problem, we measured the intensity of the focused beam as a function of the helium background gas pressure in the room temperature chamber above our diffusion pump. We varied the pressure by leaking in an added helium gas load to the chamber. We fit our results to a modified Beer's Law, which is arrived at by integrating from the skimmer to the focus, and assuming pressure is roughly constant, which we validated with a simulation in MolFlow+ \cite{molflow}. Our modified Beer's Law is then  

\begin{equation}
    S=S_0 e^{P_{He} \beta}
\end{equation}

\noindent where we have collected constants into the term $\beta$. Results of our measurements and fit are shown in Figure \ref{fig:vacuum}. This data shows that even at our lowest operating flow rate of 50 sccm, about 60\% of the lithium atoms are lost. Thus, the flux of our source would be about 2.5 times higher if we could reduce the helium pressure to much lower values. Further, if we could maintain this low vacuum pressure while also ramping up our helium flow to 150 sccm, we estimate that our source would achieve 10 times more flux than the maximum we've observed so far, \textit{i.e} a flux of $2 \times 10^{13}$ atoms/s. We plan to implement an improved 
vacuum system design which should allow these higher flux levels to be realized. 

\section{Conclusion}

In this work, we have described an intense cold atom $^7$Li source based on post-nozzle seeding of a continuous, cryogenic helium jet. Seeded atoms were extracted from the helium with magnetic focusing, allowing us to capture an appreciable solid angle from the source. Magnetic focusing and deflection would also allow us to effectively separate the lithium beam from the helium jet for further application in a high vacuum environment. The extracted lithium atoms have transverse temperatures below 20 mK, and a longitudinal temperature of 7(3) mK, substantially colder than previous seeded jet sources. Our beam velocity of 210 m/s was relatively low, due to the low temperature of our jet source. We realized an extracted beam flux of about $2\times10^{12}$ atoms/s, similar to the highest flux obtained with previous cold atom beam sources \cite{Foot_2021_pyramid_mot}. This flux was limited by the loss of $^7$Li atoms from collisions with helium background gas, and we estimate that we could produce a 10 times higher flux with a modified vacuum system design.

We have modeled the capture and focusing of Li atoms by our helium jet. This provided us with a quantitiative understanding of our focused beam diameter and seeding efficiency, and can also provide guidance  for further improvements in the source. Besides the vacuum modifications, it should be possible to further optimize this source with modifications to the nozzle shape, Li source, skimmer, and lens. It would also be of interest to combine this method with other techniques for further slowing and cooling of the beam, including laser-cooling. 

$^7$Li atoms were seeded in this work, but this method should be adaptable to other paramagnetic atoms and molecules. Cold atomic hydrogen or metastable helium beams could be produced. It would also be of interest to explore the suitability of this method for cooling of much more massive species, and for ro-vibrational cooling of molecules. For instance, a heavy paramagnetic molecule such as YbF is of interest in searches for time-reversal symmetry violation \cite{Hudson_2011_YbF_edm}. 

Some potential applications of intense atomic beams, such as loading atomic guides with sufficient density for evaporative cooling \cite{Lahaye_2005_evaporation_in_guide, Olson_2006_cold_beam_in_guide}, or atomic holography  \cite{shimizu_2000_atom_holography}, have up to now been substantially limited by the brightness of available atomic beam sources. With further advances in source brightness, these applications may become more practical. Our source should be particularly well-adapted to pumping a neutral atom waveguide, since our beam focus is produced by a magnetic lens and therefore has a phase-space distribution that can be readily accepted by a magnetic guide. This beam source could also prove useful in other applications, such as cold collision studies and precision measurements.  

\section{Acknowledgements}

We acknowledge Eite Tiesinga and Jacek K\l{}os, who shared their calculated He-Li cross sections with us, and useful contributions to this experiment by Collin Diver, Brent Kurtzel, Christian Gage Brandt, Grant Gorman, Kevin Wen, Brady Stoll, and Travis Briles. 

We gratefully acknowledge the financial support of this work by Fondren Foundation and the Army Research Laboratory Cooperative Research and Development Agreement number 16-080-004.

\bibliography{references}

\end{document}